\documentclass{elsart}
\usepackage{graphicx}
\usepackage{amssymb}
\usepackage{rotating}
\usepackage{multirow}

\newcommand{\mygraphics}[1]{%
  \includegraphics[width=\linewidth,clip=true]{#1}\vspace{0pt} 
}
\newcommand{\mygraphscl}[2]{%
  \includegraphics[width=#2\linewidth,clip=true]{#1}\vspace{0pt} 
}

\def\IEEEtns                 {\mbox{IEEE Trans. on Nucl. Sci.}}

\def\NIM                     {\mbox{Nucl. Instrum. and Methods}}

\def\ZPhys                   {\mbox{Z. Phys.}}
\newcommand{\Journal}[4]     {{#1} {\bf {#2}}, {#3} ({#4})}

\newcommand{\WWWAddr}[1]     {{\tt {#1}}}
\newcommand{\FermiPub}[2]    {Fermilab-Pub-{#1}/{#2}}

\newcommand{\ChipRef}[4]     {{#1}, {#2} ({#3})\\ \WWWAddr{#4}}

\newcommand {\unitexp}[2] {\mbox{{#1}$^{\mathrm{#2}}$}}
\newcommand {\scinot}[2]  {\mbox{{#1}$\times$10$^{\mathrm{#2}}$}}

\newcommand {\XxY}[2]     {\mbox{{#1}$\times${#2}}}

\newcommand {\ra}       {\mbox{$\rightarrow$}}

\newcommand {\Dzero}    {\mbox{D0}}

\newcommand{\ie}         {\mbox{i.e.}}
\newcommand{\eg}         {\mbox{e.g.}}

\newcommand {\TeV}    {\mbox{TeV}}
\newcommand {\GeV}    {\mbox{GeV}}

\newcommand {\us}     {\mbox{$\mu$s}}
\newcommand {\instLunit} {\unitexp{cm}{-2}\unitexp{s}{-1}}
\newcommand {\Ohm}    {\mbox{$\Omega$}}

\newcommand {\EMtt}      {\mbox{EM}}
\newcommand {\HADtt}     {\mbox{HD}}
\newcommand {\TTC}       {\mbox{TTCL}}
\newcommand {\abseta}    {\mbox{$\mid\eta\mid$}}
\newcommand {\deta}      {\mbox{$\Delta \eta$}}
\newcommand {\dphi}      {\mbox{$\Delta \phi$}}
\newcommand {\etaxphi}   {\mbox{$\eta \times \phi$}}
\newcommand {\detaxdphi} {\mbox{$\Delta\eta \times \Delta\phi$}}
\newcommand {\Et}        {\mbox{$E_T$}}
\newcommand {\Ex}        {\mbox{$E_x$}}
\newcommand {\Ey}        {\mbox{$E_y$}}

\newcommand {\Zee}        {\mbox{$Z \ra e^+ e^-$}}
\newcommand {\Ztautau}    {\mbox{$Z \ra \tau^+ \tau^-$}}

\newcommand {\init}      {\mbox{\tt INIT}}
\newcommand {\firstturn} {\mbox{\tt TURN}}
\newcommand {\clk}       {\mbox{\tt CLK7}}
\newcommand {\realbx}    {\mbox{\tt REALBX}}
\newcommand {\accept}    {\mbox{\tt L1ACCEPT}}
\newcommand {\error}     {\mbox{\tt L1ERROR}}
\newcommand {\busy}      {\mbox{\tt L1BUSY}}
\newcommand {\bxno}      {\mbox{\tt BX\_NO}}
\newcommand {\monitor}   {\mbox{\tt MONITOR}}

\newcommand {\tabrun}    {\mbox{\tt TAB\_RUN}}
\newcommand {\tabtrig}   {\mbox{\tt TAB\_TRIG}}
\newcommand {\tabfrm}    {\mbox{\tt TAB\_FRM}}
\newcommand {\tabaddr}   {\mbox{\tt TAB\_ADDR}}
\newcommand {\tabdata}   {\mbox{\tt TAB\_DATA}}

\newcommand {\adfmon}    {\mbox{\tt ADF\_MON}}
\newcommand {\adftrig}   {\mbox{\tt ADF\_TRIG}}

\begin{document}

%==================================================================
% TITLE PAGE
%==================================================================
\begin{frontmatter}

% only for Fermilab and ArXiv preprints
%\begin{flushright}
%  Fermilab-Pub-07/492-E
%\end{flushright}

\title{ 
  Design and Implementation of the New D0 Level-1 Calorimeter Trigger 
}

%--------------------------
% Author List
%--------------------------
\author[msu]{M.~Abolins}
\author[uic]{M.~Adams}
\author[fsu]{T.~Adams}
\author[alberta,york]{E.~Aguilo}
\author[fnal]{J.~Anderson}
\author[fnal]{L.~Bagby}
\author[cu]{J.~Ban}
\author[neu]{E.~Barberis}
\author[york]{S.~Beale}
\author[msu]{J.~Benitez}
\author[msu]{J.~Biehl}
\author[fnal]{M.~Bowden}
\author[msu]{R.~Brock}
\author[saclay]{J.~Bystricky}
\author[dublin]{M.~Cwiok}
\author[saclay]{D.~Calvet}
\author[fnal]{S.~Cihangir}
\author[msu]{D.~Edmunds}
\author[iu]{H.~Evans}
\author[neu]{C.~Fantasia}
\author[fnal]{J.~Foglesong}
\author[fnal]{J.~Green}
\author[cu]{C.~Johnson}
\author[smu]{R.~Kehoe}
\author[cu]{S.~Lammers}
\author[msu]{P.~Laurens}
\author[saclay]{P.~Le~D\^{u}}
\author[iu,cppm]{P.S.~Mangeard}
\author[cu]{J.~Mitrevski}
\author[cu]{M.~Mulhearn}
\author[saclay]{M.~Mur}
\author[delhi,now-fnal]{Md.~Naimuddin}
\author[cu]{J.~Parsons}
\author[rice]{G.~Pawloski}
\author[saclay]{E.~Perez}
\author[smu]{P.~Renkel}
\author[neu]{A.~Roe}
\author[cu]{W.~Sippach}
\author[uic,now-fnal]{A.~Stone}
\author[york]{W.~Taylor}
\author[msu]{R.~Unalan}
\author[uic]{N.~Varelas}
\author[fnal]{M.~Verzocchi}
\author[msu]{H.~Weerts}
\author[neu]{D.~R.~Wood}
\author[cu]{L.~Zhang}
\author[fnal]{T.~Zmuda}

%\address[argonne] {Argonne National Laboratory,
%                   Argonne, Illinois 60439, USA}
\address[cu]      {Columbia University, 
                   New York, New York 10027, USA}
\address[saclay]  {DAPNIA/Service de Physique des Particules, CEA,
                   Saclay, France}
\address[delhi]   {Delhi University,
                   Delhi, India}
\address[fnal]    {Fermi National Accelerator Laboratory, 
                   Batavia, Illinois 60510, USA}
\address[fsu]     {Florida State University, 
                   Tallahassee, Florida 32306, USA}
\address[iu]      {Indiana University,
                   Bloomington, Indiana 47405 USA}
\address[msu]     {Michigan State University,
                   East Lansing, Michigan 48824, USA}
\address[neu]     {Northeastern University,
                   Boston, Massachusetts 02215, USA}
\address[rice]    {Rice University,
                   Houston, Texas 77005, USA}
\address[smu]     {Southern Methodist University,
                   Dallas, Texas 75275, USA}
\address[dublin]  {University College Dublin,
                   Dublin, Ireland}
\address[alberta] {University of Alberta,
                   Edmonton, Alberta, Canada}
\address[uic]     {University of Illinois at Chicago,
                   Chicago, Illinois 60607, USA}
\address[york]    {York University,
                   Toronto, Ontario, Canada}

\address[now-fnal] {Now at Fermi National Accelerator Laboratory, 
                   Batavia, Illinois 60510, USA}

\address[cppm]     {Now at Universit\'{e} d'Aix, 
                   Centre de Physique des Particules de Marseille,
                   Marseille, France}

%--------------------------
% Abstract
%--------------------------
\begin{abstract}
Increasing luminosity at the Fermilab Tevatron collider has led the
\Dzero\ collaboration to make improvements to its detector beyond
those already in place for Run IIa, which began in March 2001.
One of the cornerstones of this Run IIb upgrade is a completely
redesigned level-1 calorimeter trigger system.
The new system employs novel architecture and algorithms to
retain high efficiency for interesting events while substantially
increasing rejection of background.
We describe the design and implementation of the new level-1 calorimeter
trigger hardware and discuss its performance
during Run IIb data taking.
In addition to strengthening the physics capabilities of \Dzero , this
trigger system will provide valuable insight into the operation of
analogous devices to be used at LHC experiments.
\end{abstract}

%--------------------------
% Keywords and PACs
%--------------------------

\end{frontmatter}

%==================================================================
% INTRODUCTION
%==================================================================
\section{Introduction}
\label{sect:intro}
During the five year period between the end of Run I in 1996 and
the beginning of Run IIa in 2001, the Fermilab Tevatron accelerator
implemented an ambitious upgrade program \cite{run2tev} in which the
proton--antiproton center of mass energy was increased from 1.8 TeV to
1.96 \TeV\ 
and the instantaneous luminosity was boosted by an order of
magnitude. 
To take advantage of the new accelerator conditions,
the collider experiments, CDF and \Dzero , 
also embarked on major upgrades to their
detectors. 

The \Dzero\ upgrade, described fully in \cite{d0run2},
involved a complete replacement of the Run I tracking system with a
new set of silicon micro-strip and scintillating fiber trackers as
well as the addition of a 2 T solenoid magnet. 
Although the uranium and liquid argon calorimeter was left unchanged, its
electronics were overhauled to match the new Tevatron bunch structure,
and a series of preshower detectors was added outside of the 
solenoid to help measure energy of electrons, photons, and jets.
Muon detection was improved with the addition of new detectors and
shielding. 
Finally, the trigger and data
acquisition systems were almost completely redesigned.

As originally proposed \cite{run2tev}, approximately 20 times the
integrated luminosity delivered in Run I
was scheduled to be accumulated during Run II,
for a total of 2 \unitexp{fb}{-1}. 
To accomplish this
goal major improvements were made to all aspects of the Tevatron,
particularly in the areas of antiproton production.
The bunch structure of the machine was also changed to accommodate
\XxY{36}{36} bunches of \XxY{protons}{antiprotons}, with an inter-bunch
spacing of 396 ns, which is 
an improvement over the \XxY{6}{6} mode of operation in Run I.
Future enhancements to 132 ns inter-bunch spacing were also foreseen,
motivating a Tevatron RF structure
with 159 potential bunch crossings (separated by 132 ns) 
during the time it
takes a proton or antiproton to make a single revolution, or
{\it turn} around the Tevatron. 
Of these potential crossings, only 36
contain actual proton--antiproton collisions.

Driven by ambitious physics goals of the experiments, 
a series of continued Tevatron improvements
was also planned \cite{run2tevupgr},
beyond the Run II baseline,
with the aim of increasing the total integrated luminosity collected
to the 4 -- 8 \unitexp{fb}{-1} level.
To achieve this performance, instantaneous luminosities in excess of
\scinot{2}{32} \instLunit\ are required.
Tevatron upgrades for this period include
fully commissioning the Recycler
as a second stage of antiproton storage
and implementing electron cooling in the Recycler.
The majority of these improvements were successfully completed during
a Tevatron shutdown lasting from February to May 2006,
which marks the beginning of Run IIb.

The long-term effects of the Run IIb Tevatron upgrade on the
\Dzero\ 
experiment are threefold. First, the additional 
integrated luminosity 
to be delivered to \Dzero\ during the course of Run IIb
will also increase the total radiation
dose accumulated by the silicon detector.
Best estimates indicate that such a dose will compromise the
performance of the inner layer of the detector,
affecting the ability of \Dzero\ to tag $b$-quarks --
a necessary
ingredient in much of the experiment's physics program. 
Second, the increased instantaneous luminosity stresses the trigger
system, decreasing the ability to reject background while maintaining
high efficiency for signal events.
And finally, the plan of having real bunch crossings separated
by 132 ns,
although not realized in the final Run IIb configuration, 
would have
created problems matching calorimeter signals with
their 
correct bunch crossing in the Run IIa calorimeter trigger system.

The first of the effects mentioned above led \Dzero\ to propose the
addition of a
radiation-hard inner silicon layer ({\it Layer-0}) to the
tracking system \cite{layer0}.
The second and third effects required changes to various aspects of
the trigger system \cite{trig-2b}. 
These additions and modifications, 
collectively referred to as the {\it \Dzero\ Run IIb Upgrade},
were designed and implemented between 2002 and 2006 and were installed
in the experiment during the 2006 Tevatron shutdown.

In the following we describe the Level-1 Calorimeter
Trigger System (L1Cal) designed for operation during Run IIb.
Section \ref{sect:existing} contains a brief description of the
Run IIa \Dzero\ calorimeter and the three-level trigger system. 
Section \ref{sect:motivation} discusses the motivation for replacing the
L1Cal trigger, which was used in Run I and Run IIa.
Algorithms used in the new system and their simulation are described
in Sections \ref{sect:algos} and \ref{sect:sim},
while the hardware designed to implement these algorithms is
detailed in Sections \ref{sect:hardware}, \ref{sect:adfsyst}, 
\ref{sect:adf-to-tab}, and \ref{sect:tab-gab}.
Mechanisms for online control and monitoring of the new L1Cal are
outlined in Sections \ref{sect:online} and \ref{sect:monitor}.
This article then concludes with a discussion of early calibration and
performance results in Sections \ref{sect:calib} and
\ref{sect:results}, with a summary presented in Section
\ref{sect:conclusion}.

%==================================================================
% EXISTING FRAMEWORK
%==================================================================
\section{Existing Framework}
\label{sect:existing}

\subsection{The \Dzero\ Calorimeter}
\label{sect:cal2a}
\begin{figure}
\begin{minipage}[t]{0.475\textwidth}
  \begin{center}
    \mygraphics{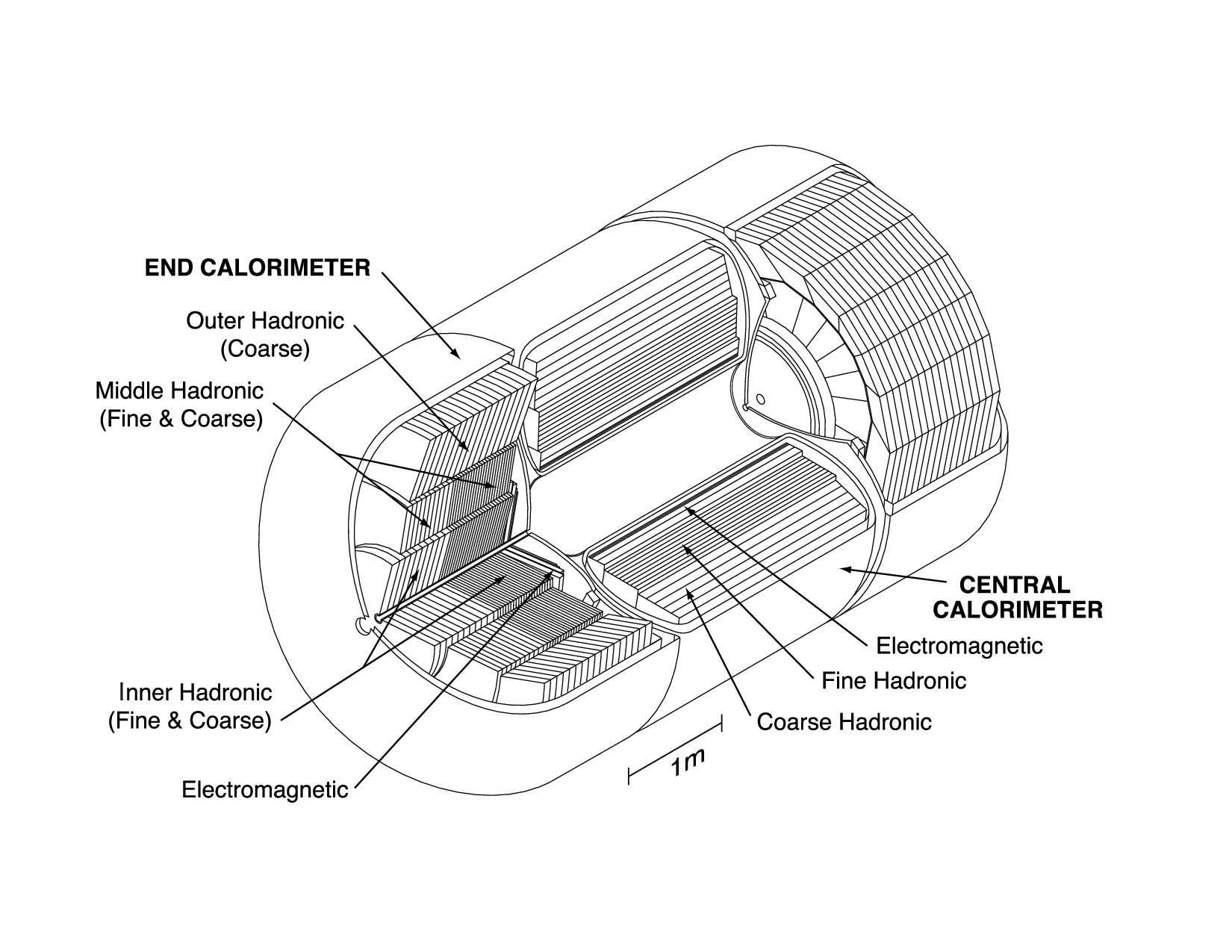}
  \end{center}
\end{minipage}
\hfill
\begin{minipage}[t]{0.475\textwidth}
  \begin{center}
    \mygraphics{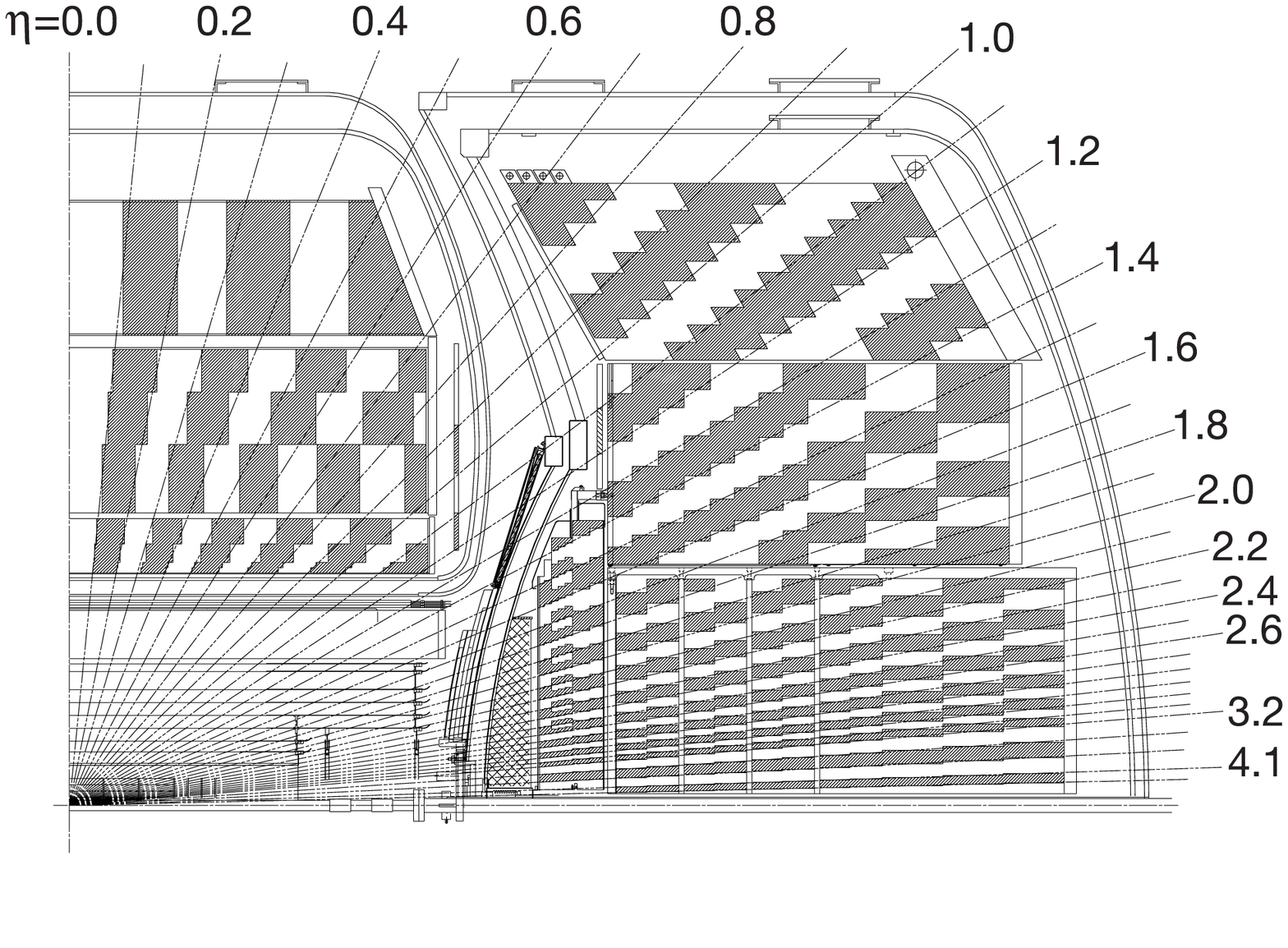}
  \end{center}
\end{minipage}
\begin{center}
  \caption{An isometric view of the central and two endcap
           calorimeters (left) and
	   a schematic view of a portion of the calorimeter
	   showing the transverse and longitudinal segmentation
           pattern (right).}
  \label{fig:d0cal} 
\end{center}
\end{figure}

The basis of the Run IIb L1Cal trigger is the \Dzero\ calorimeter,
described in more detail in \cite{d0run2,d0run1}. This detector, shown
schematically in Fig.\ \ref{fig:d0cal}, consists of three sampling
calorimeters (a barrel and two endcaps),
in three separate cryostats, 
using liquid argon as the active medium and depleted
uranium, uranium-niobium alloy, copper or stainless steel as the
absorber. It also includes detectors in the intercryostat region (ICR), 
where the barrel and endcaps meet,
consisting of scintillating tiles, as well as instrumented regions of the
liquid argon without absorbers.
The calorimeter has three longitudinal sections --
electromagnetic (EM), fine hadronic (FH) and coarse hadronic (CH) --
each themselves divided into several layers.
It is segmented laterally into cells of size
\detaxdphi\ $\sim$ \XxY{0.1}{0.1} \cite{etaphi}
arranged in pseudo-projective towers
(except for one layer in the EM section, which has
\detaxdphi $\sim$ \XxY{0.05}{0.05}).
The calorimeter system provides coverage out to 
\abseta\ $\sim$ 4.

% Fig.s 30,31 from D0 NIM

Charge collected in the calorimeter is transmitted via
impedance-matched coaxial cables of $\sim$10 m length to charge sensitive
preamplifiers located on the detector. The charge integrated output of
these preamplifiers has a rise time of $\sim$450 ns, corresponding to
the electron drift time across a liquid-argon gap, and a fall time of
$\sim$15 \us . The single-ended preamplifier signals are sent over
$\sim$25 m of twisted pair cable to Baseline Subtractor (BLS) cards.

%Fig. 34 from D0 NIM
\begin{figure}
\begin{center}
  \mygraphics{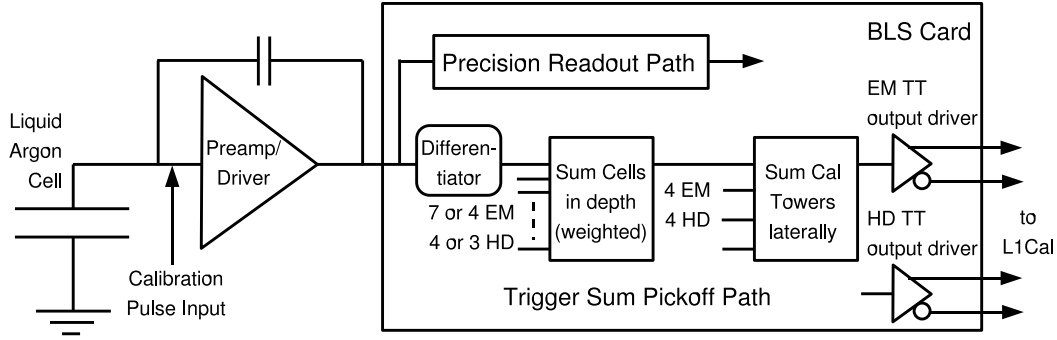}
  \caption{The calorimeter readout chain, including a preamplifier and
           baseline subtractor card (BLS) with emphasis on the
           elements of the trigger sum pickoff path.}
  \label{fig:bls} 
\end{center}
\end{figure}

On the 1152 BLS cards, the preamplifier signals are split into two paths:
the {\it precision readout} and the {\it trigger sum pickoff}.
Precision readout path signals for each calorimeter cell are shaped,
baseline subtracted, and
stored in a set of switched capacitor arrays 
awaiting Level-1 and Level-2 trigger decisions. 
Signals on the trigger sum pickoff path,
shown in Fig.\ \ref{fig:bls}
are shaped to a triangular pulse with a fast rise and a
linear fall over 400 ns. 
They are then passed to analog summers that 
add signals in different cells, 
weighted appropriately for the sampling fraction and capacitance
of each cell 
to form \EMtt\ and \HADtt\ trigger towers (TT). 
\EMtt\ TTs
contain all cells (typically 28) in \XxY{0.2}{0.2} \detaxdphi\ regions
of the EM section 
of the calorimeter, while \HADtt\ TTs use (typically 12) cells in the
FH section of the 
calorimeter to form \XxY{0.2}{0.2} regions. This granularity leads to
1280 \EMtt\ and 1280 \HADtt\ TTs forming a \XxY{40}{32} grid in \etaxphi\
space, which covers the entire azimuthal region for \abseta\ $<$
4.0. 
Due mainly to overlapping collisions,
which complicate the forward environment, 
however, only the region
\abseta\ $<$ 3.2 is used for triggering.

\begin{figure}
\begin{minipage}[t]{0.475\textwidth}
  \begin{center}
    \mygraphics{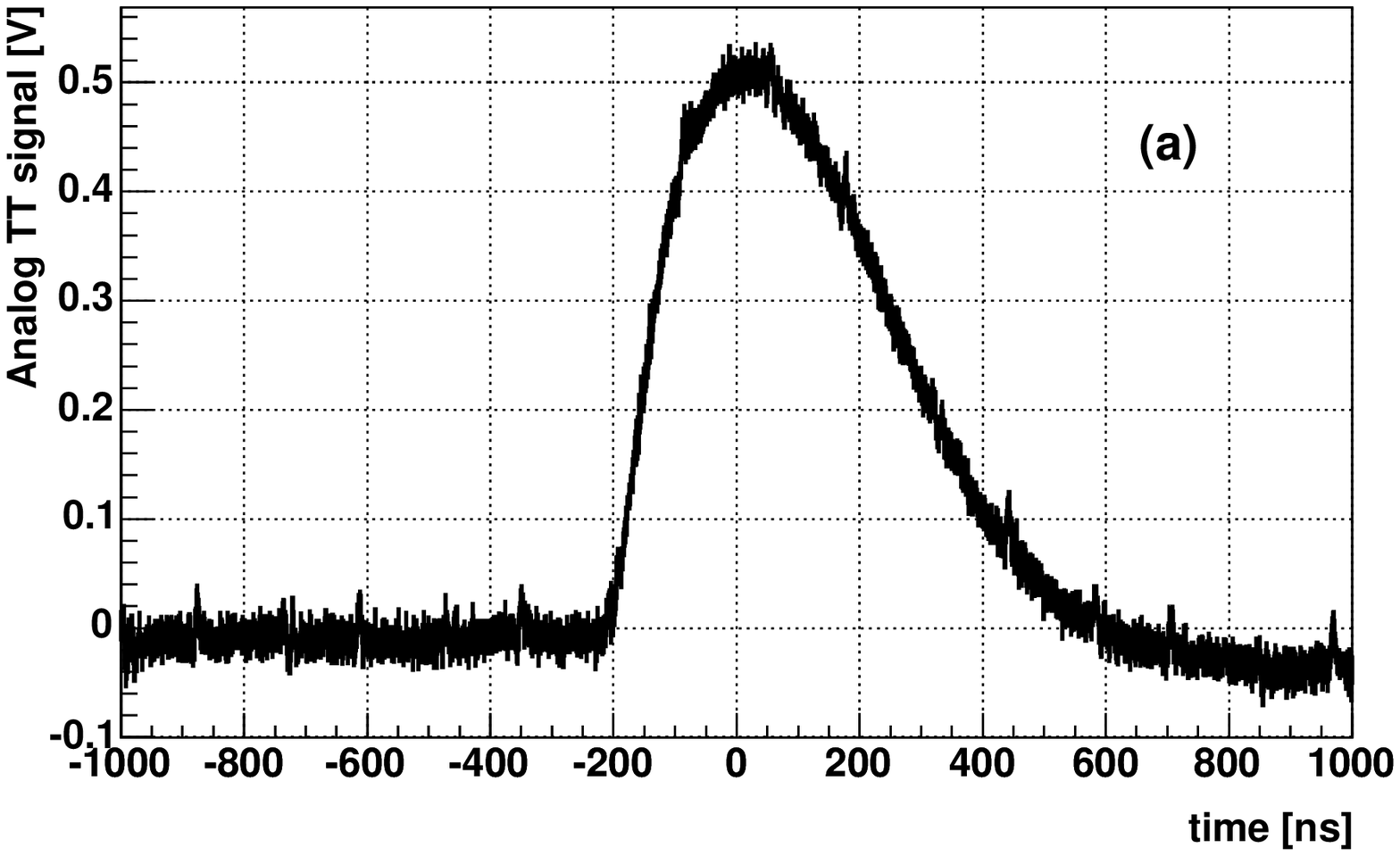}
  \end{center}
\end{minipage}
\hfill
\begin{minipage}[t]{0.475\textwidth}
  \begin{center}
    \mygraphics{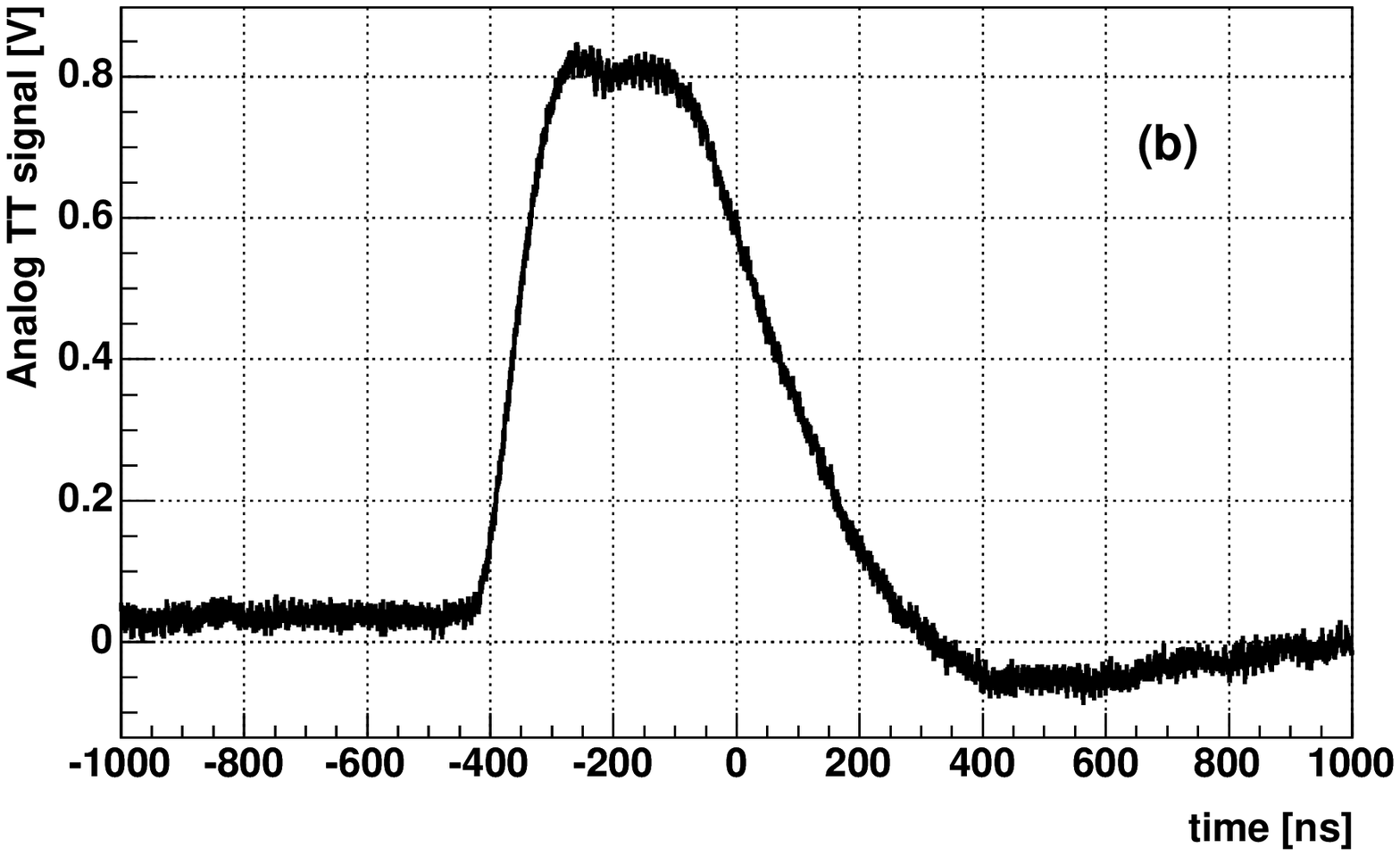}
  \end{center}
\end{minipage}
\begin{center}
  \caption{Typical \EMtt\ (a) and \HADtt\ (b) analog signals. 
           In both plots, the
           non-inverted minus the inverted
           differential signals are shown.}
  \label{fig:ttsignal} 
\end{center}
\end{figure}

The \EMtt\ and \HADtt\ TT signals are transmitted differentially to the
Level-1 Calorimeter Trigger electronics on two separate miniature
coaxial cables. Although the signal characteristics of these cables
are quite good, some degradation occurs in the transmission,
yielding L1Cal input signals with a rise time of
$\sim$250 ns and a total duration of up to 700 ns. Typical \EMtt\ and
\HADtt\ TT signals are shown in Fig.\ \ref{fig:ttsignal}.

\subsection{Overview of the \Dzero\ Trigger System}
\label{sect:trigover}
\begin{figure}
\begin{center}
  \mygraphics{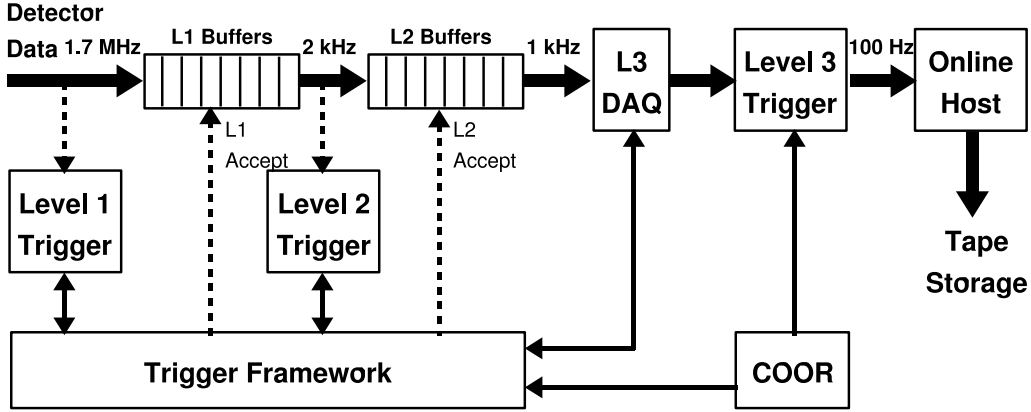}
  \caption{An overview of the \Dzero\ trigger and data acquisition
           system.} 
  \label{fig:d0trig} 
\end{center}
\end{figure}

\begin{figure}
\begin{center}
  \mygraphscl{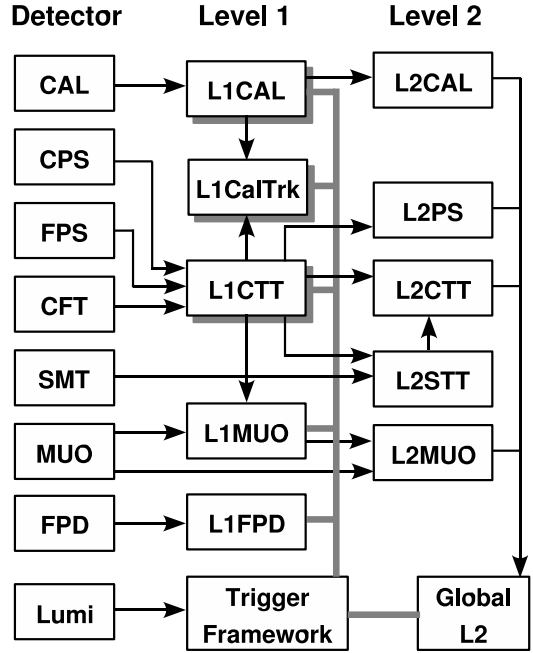}{0.5}
  \caption{A block diagram of the \Dzero\ L1 and L2 trigger
           systems.}
  \label{fig:l1l2-2b} 
\end{center}
\end{figure}

The \Dzero\ experiment uses a three level trigger system,
shown schematically in Fig.\ \ref{fig:d0trig}
and described in more detail in \cite{d0run2},
to select
interesting events from the 1.7 MHz of bunch crossings seen in the
detector. 
Individual elements contributing to the Level-1 (L1) and Level-2 (L2)
systems, as used in Run IIb,
are shown in Fig.\ \ref{fig:l1l2-2b}.

The L1 trigger system,
implemented in custom hardware,
examines data from the detector for every bunch
crossing. It consists of separate elements for calorimeter (L1Cal),
scintillating fiber tracking (L1CTT), 
muon (L1Muon), and forward proton (L1FPD) data. New
for Run IIb is an element that matches tracks and calorimeter clusters
at L1 (L1CalTrk), which is functionally similar to L1Muon.
Each L1 trigger element sends its decisions on a set of criteria (for
example, the presence of two jets with transverse energy above a
threshold) to the trigger framework (TFW).
The TFW uses these decisions, referred to as the {\it and/or} terms to
decide whether the event should be accepted for further processing or
rejected. 
Because of the depth of data pipelines in the detector's front end
electronics, L1 decisions from each of the trigger elements must
arrive at the TFW within 3.7 \us\ of the bunch crossing
producing their data. 
This pipeline depth was increased from its Run IIa value of 3.3 \us\ 
in order to accomodate the extra latency induced
by the L1CalTrk system.
After an L1 accept, data are transferred off of the pipelines, 
inducing deadtime in the system.
The maximum allowable L1 accept rate,
generally around 2 kHz,
is set by the desire to limit this deadtime to the 5\% level.

The L2 system receives data from the detector and from the L1 trigger
elements on each L1 accept. It consists of detector-specific
pre-processor engines for calorimeter (L2Cal); preshower (L2PS);
scintillating fiber (L2CTT) and silicon (L2STT) tracking; and muon
(L2Muon) data. Processed data from each of these elements is
transmitted to a global processor (L2Global) that selects events based
on detector-wide correlations between its input elements. The L2 trigger
operates at a maximum input rate of 2 kHz and
provides L2 accepts at a rate of up to 1 kHz.

The final stage in the \Dzero\ trigger system, Level-3 (L3), 
consists of a farm
of PCs that have access to the full detector readout on L2
accepts. These processors run a simplified version of the offline
event reconstruction and make decisions based on physics objects and
the relationships between them. L3 accepts events for permanent
storage at a rate of up to 150 Hz (typically, 100 Hz).

The configuration of the entire \Dzero\ trigger system is accomplished
under the direction of the central coordination program (COOR), which
is also used for detector configuration and run control.

%==================================================================
% MOTIVATION
%==================================================================
\section{Motivation for the L1Cal Upgrade}
\label{sect:motivation}
By the time of the start of Run IIa in 2001, there was
already a tentative plan in place for an extension to the run with
accompanying upgrades to the accelerator complex \cite{run2tevupgr},
leading to an additional 2--6 \unitexp{fb}{-1} of integrated luminosity
beyond the original goal of 2 \unitexp{fb}{-1}.
This large increase in statistical power opens new possibilities
for physics at the Tevatron such as greater precision in critical
measurements like the top quark mass and W boson mass, the possibility of
detecting or excluding very rare Standard Model processes (including
production of the Higgs boson), and greater sensitivity for beyond
the Standard Model processes like supersymmetry.

At a hadron collider like the Tevatron, however, only a small
fraction of the collisions can be recorded, and it is the trigger
that dictates what physics processes can be studied and what is
left unexplored. The trigger for the \Dzero\ experiment in Run IIa had
been designed for a maximum luminosity of 
\scinot{1}{32} \instLunit , while the
peak luminosities in Run IIb are expected to go as high as 
\scinot{3}{32} \instLunit .
In the three-level trigger system employed by \Dzero , only the
L3 trigger can be modified to increase its throughput; the
maximum output rates at L1 and L2 are imposed by
fundamental features of the subdetector electronics.  
Thus, fitting L1 and L2 triggers into the bandwidth limitations of the
system can only be accomplished by increasing their rejection power.
While an increase in the transverse energy thresholds
at L1 would have been a simple way to achieve
higher rejection, such a threshold increase would be too costly in
efficiency for the physics processes of interest.
For example, raising the thresholds on the two jet triggers used in
the search for $p \bar{p} \ra Z H \ra \nu \bar{\nu} b \bar{b}$ events
to yield acceptable rates would have resulted in a decrease in signal
efficiency of more than 20\%.
The
\Dzero\ Run IIb Trigger Upgrade \cite{trig-2b} was
designed to achieve the necessary rate reduction through
greater selectivity, particularly at the level of individual L1
trigger elements.

The L1Cal trigger used in Run I and in Run IIa \cite{l1cal-2a}
was based on counting individual trigger towers above thresholds in
transverse energy (\Et ).  Because the energy from electrons/photons
and especially
from jets tends to spread over multiple TTs, the
thresholds on tower \Et\ had to be set low relative to the desired
electron or jet \Et . For example, an \EMtt\ trigger tower
threshold of 5 \GeV\ is fully efficient only for electrons with \Et\
greater than about 10 \GeV , and a 5 \GeV\ threshold for \EMtt +\HADtt\
tower \Et\ only becomes 90\% efficient for jet transverse energies
above 50 \GeV .

The primary strategy of the Run IIb upgrade of L1Cal is therefore
to improve the sharpness of the thresholds for electrons, photons
and jets by forming clusters of TTs and comparing
the transverse energies of these clusters, rather than 
individual tower \Et s, to thresholds.  

The design of clustering using sliding windows
(see Section \ref{sect:algos})
in Field Programmable Gate Arrays (FPGAs) 
meets the requirements of this strategy, and also opens
new possibilities for L1Cal, including sophisticated
use of shower shape and isolation; algorithms to find hadronic
decays of tau leptons through their characteristic transverse profile;
and requirements on the topology of the electrons,
jets, taus, and missing transverse energy in an event.

%==================================================================
% ALGORITHMS
%==================================================================
\section{Algorithms for the Run IIb L1Cal}
\label{sect:algos}
Clustering of individual TTs into EM and Jet objects is accomplished
in the Run IIb L1Cal by the use of a sliding windows (SW) algorithm.
This algorithm performs a highly parallel cluster search in
which groups of contiguous TTs are compared to nearby groups to
determine the location of local maxima in \Et\ deposition.
Variants of the SW algorithm have been studied
extensively at different HEP experiments \cite{swstudy}, 
and have been found to be
highly efficient at triggering on EM and Jet objects, while not having
the latency drawbacks of iterative clustering algorithms.  For a full
discussion of the merits of the sliding
windows algorithm, see \cite{run2b-tdr}.

\begin{figure}[h] 
\begin{center}
  \mygraphics{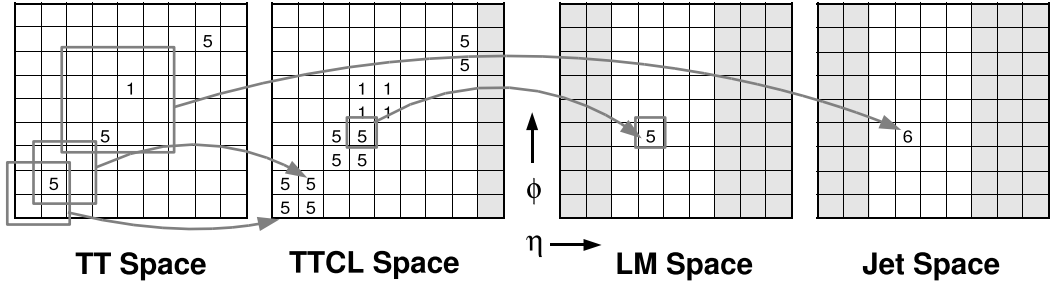}
  \caption{
    The stages of algorithm flow for the sliding windows algorithm.
    In this example, which corresponds to the Run IIb Jet algorithm, 
    a \XxY{2}{2} TT \TTC\ is used, 
    indexed by the position of its smallest $\eta ,\phi$ TT.
    Baseline subtracted TT energies are indicated by numbers, and
    local maxima are required to be separated
    by at least 1 TT. Jet objects are defined as the \Et\ sum of the
    \XxY{4}{4} TTs centered on the \TTC .
    Light gray regions in the diagrams indicate areas for which the
    object in question cannot be constructed because of boundary
    effects.}
  \label{fig:lm_finding}
\end{center}
\end{figure}

The implementation of the sliding windows algorithm in the \Dzero\
calorimeter trigger occurs in three phases.  In the first phase, 
the digitized transverse energies of several
TTs are summed into Trigger Tower Clusters (\TTC ). 
These \TTC\ sums, based on the size
of the EM or Jet sliding window, are constructed for every point 
in trigger tower space, and are indexed by the $\eta ,\phi$ coordinate
of one of the contributing TTs,
with different conventions being used for different algorithms
(see Sections \ref{sect:jetalgo} and \ref{sect:emalgo}).
This process, which yields a grid of
\TTC s that share energy with their close neighbors, is shown in the
first and second panels of Fig.\ \ref{fig:lm_finding}.
  
In the second phase, the \TTC s are analyzed to
determine locations of large energy deposits called local maxima
(LM).  These LM are chosen based on a comparison of the magnitude of
the \Et\ of
a \TTC\ with that of its adjacent \TTC s.  
Multiple
counting of Jet or EM objects is avoided by requiring a spatial
separation between adjacent local maxima as illustrated in the third
panel of Fig.\ \ref{fig:lm_finding}.

In the third phase,
additional information is added to define an output object.
In the case of Jet objects, shown in the fourth panel of Fig.\
\ref{fig:lm_finding}, energy of surrounding TTs is added to the \TTC\
energy to give the total Jet object energy.
EM and Tau objects are also refined in this phase using isolation
information (see Sections \ref{sect:emalgo} and \ref{sect:taualgo}).

Results for the entire calorimeter can be obtained very quickly using
this type of algorithm 
by performing the LM finding and object refinement phases of the
algorithm in parallel for each \TTC .

%----------------------------------------------------------------
\subsection{Jets}
\label{sect:jetalgo}
Jets at the Tevatron have lateral sizes of order one unit in 
$\eta ,\phi$ space 
and deposit energy in both the
electromagnetic and hadronic portions of the calorimeter.  
Therefore,
Jet objects in the \Dzero\ L1Cal are defined using the sum of the
\EMtt\ and \HADtt\ energies as the input to the \TTC -sums.
The \TTC s are \XxY{2}{2} in trigger tower units, corresponding to a region
\XxY{0.4}{0.4} in \etaxphi\ space on the inner face of the calorimeter.
Local maxima are required to be separated by one trigger tower and the
final energy sums are \XxY{4}{4} in TT space, corresponding to a
region \XxY{0.8}{0.8} in \etaxphi\ space.

The values of these clustering parameters were determined by
optimizing Jet object energy and position resolution.

%-------------------------------------------------------------------
\subsection{EM Objects}
\label{sect:emalgo}
\begin{figure}[h]
\begin{center}
  \mygraphscl{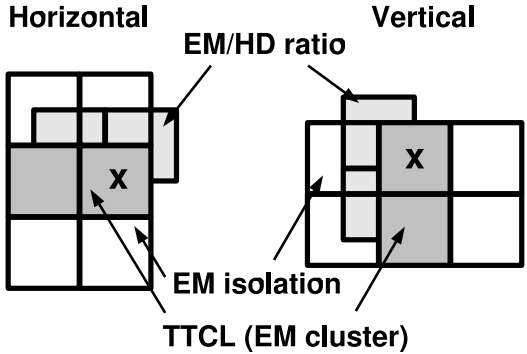}{0.5}
  \caption{Definition of EM trigger objects.}
  \label{fig:emobj}
\end{center}
\end{figure}

EM objects (electrons or photons)
have lateral shower profiles that are much smaller than the TT size,
and tend not to deposit energy in the hadronic calorimeter.  
For this reason, \EMtt\ TTs are input directly
to the local maximum finding algorithm (the \TTC\ size is \XxY{1}{1} in
TT units).
Because electrons or photons may deposit energy close to the boundary
between TTs,
the final EM object,
as shown in Fig.\ \ref{fig:emobj},
is comprised of two adjacent trigger towers, 
oriented horizontally (containing two TTs in $\eta$) 
or vertically (containing two TTs in $\phi$), 
where the first tower is the LM and the second is the
neighboring tower with the highest \Et .  
Cuts can also be applied on the
electromagnetic fraction (\EMtt /\HADtt ) 
and isolation of the candidate EM object.
The former is determined using the ratio of the \EMtt\
TT energies making up the EM object and the corresponding two
\HADtt\ TTs directly behind it.  The isolation region is composed of
the four \EMtt\ TTs adjacent to the EM object; cuts are placed on
the ratio of the total \Et\ in the EM-isolation region and the EM
object \Et .  In
both cases, the ratio cut value is constrained to be a power of two
in order to reduce latency in the divide operation as
implemented in digital logic.

This algorithm was chosen
based on an optimization of the efficiency for triggering on electrons
from $W\rightarrow e\nu$ and
$J/\psi \rightarrow e^+ e^-$ decays.  

%------------------------------------------------------------------
\subsection{Taus}
\label{sect:taualgo}
Tau leptons that decay hadronically look similar to jets, but
have narrow, energetic cores.  
This allows extra efficiency for processes containing taus to be
obtained by relaxing \Et\ threshold requirements on these objects
(compared to Jet thresholds) but additionally requiring that only
small amounts of energy surround the tau candidate.
The Run IIb L1Cal uses the results of the Jet algorithm as a basis for
Tau objects but also calculates the ratio of the \XxY{2}{2} TT \TTC\ to
the \XxY{4}{4} total Jet object \Et . 
Large values of this isolation ratio,
as well as large \Et , 
are required in the definition of a Tau object. 
Because of data transfer constraints in the system, however, the \Et\
associated with the Tau object is taken from the Jet object closest in
$\phi$ to the LM passing the Tau isolation cut.

%---------------------------------------------------------------------
\subsection{Sum \Et\ and missing \Et }
\label{sect:sumalgo}
Scalar and vector \Et\ sums are computed for the \EMtt +\HADtt\ TTs.
In constructing these sums, the $\eta$ range of the contributing TTs
can be restricted and
an \Et\ threshold can be applied to the TTs entering the sums
to avoid noise contamination.

%---------------------------------------------------------------------
\subsection{Use of the Intercryostat Detectors}
\label{sect:icralgo}
Object and sum energies in the Run IIb L1Cal can be configured to
include energies seen in the ICR.
Because of complicated calibrations and relatively poor resolution in
these regions, however, this option is currently not in use. 

\subsection{Topological Triggers}
\label{sect:topoalgo}
Because of its increased processing capabilities, 
the Run IIb L1Cal can require spatial correlations between some of its
objects to create {\it topological} trigger terms.
These triggers can be used to distinguish signals that have numbers of
objects identical to those observed in large backgrounds but whose
event topologies are much rarer.
An example of such a topology occurs in associated Higgs production in which
the decay $ZH\rightarrow\nu\bar{\nu} b\bar{b}$ yields two jets acoplanar 
with respect to the beam axis,
and large missing transverse energy.  Since the only visible
energy in such an event is reflected in the jets, it is difficult to
distinguish this process from the overwhelming dijet QCD background.
The Run IIb L1Cal contains a trigger that specifically selects dijet
events in which the two jets are required to be acolinear in the
transverse plane.  Other
topological triggers that have been studied are back-to-back 
(in the transverse plane)
EM object
triggers to select events containing $J/\psi$ mesons, and 
triggers that select events with jet-free 
regions of the calorimeter containing small energy deposits,
for triggering on mono-jet events.

%==================================================================
% SIMULATION AND PREDICTIONS
%==================================================================
\section{Simulation and Predictions}
\label{sect:sim}
Two independent methods of simulating the performance of the L1Cal
algorithms have been developed: a module included in the overall
\Dzero\ trigger simulation for use with Monte Carlo or real data
events ({\it TrigSim}), and
a tool developed to estimate and extrapolate trigger rates based on
real data accumulated during special low-bias runs 
({\it Trigger Rate Tool}). Both of these methods were used to develop
a new Run IIb trigger list that will collect data efficiently up
to the highest luminosities foreseen in Run IIb.

%------------------------------------------------------------------
\subsection{Monte Carlo based Simulation}
\label{sect:trigsim}
A C++ simulation of the Run IIb L1Cal trigger has been developed, as
part of 
the full \Dzero\ trigger simulation (TrigSim) -- a
single executable program that provides a standard framework for
including code that simulates each individual \Dzero\ trigger element.
This framework allows the specification of the format of the data
transferred between trigger elements, the simulation of the time
ordering of the trigger levels and the simulation of the data
transfers.
The L1Cal portion of TrigSim emulates all aspects of the L1Cal
algorithms. 
It can be run either
as part of the full \Dzero\ trigger simulation or in a stand-alone mode
on both Monte Carlo simulated data and
real \Dzero\ data,
allowing checks on hardware performance,
as well as 
estimates of signal efficiencies and background rates, 
as part of algorithm optimization.

%--------------------------------------------------------------------
\subsection{Trigger Rate Tool}
A great benefit in designing and testing the algorithms for L1Cal
in Run IIb was the availability of real collision data from Run IIa.
In every event recorded in Run IIa, the transverse energy of every trigger
tower was saved.
These energies serve as input to a stand-alone emulation
of the Run IIb algorithms (the Trigger Rate Tool) 
used to estimate rates and object-level efficiencies from
actual data.  Special data runs were taken with low tower
thresholds, and the Trigger Rate Tool was applied to 
these runs to predict the rates for any list of emulated triggers
with a proper treatment of the correlations among triggers in the
list.  The Trigger Rate Tool was also used 
to compare the Run IIa and Run IIb trigger lists
and to extrapolate rates
from the relatively low luminosities existing when the Run IIa data
was taken to the much higher values anticipated in Run IIb.
Predictions based on results obtained from this tool indicated
that the upgraded trigger would reduce the overall Level 1 rates by
about a factor of two while maintaining equal or improved efficiency
for signal processes at the highest instantaneous luminosities
foreseen in Run IIb.

%---------------------------------------------------------------------
\subsection{Predictions}
\label{sect:pred}
Predictions of the impact of the new L1Cal sliding windows
algorithms on the L1 trigger rates and efficiencies were obtained
using simulations of dijet events and various physics processes of
interest in Run IIb.
After trying different configurations that gave the same rate
as those experienced during Run IIa, the most efficient configurations
were chosen and put in an overall trigger list to check the total
rate.

\begin{figure}
\begin{center}
  \mygraphics{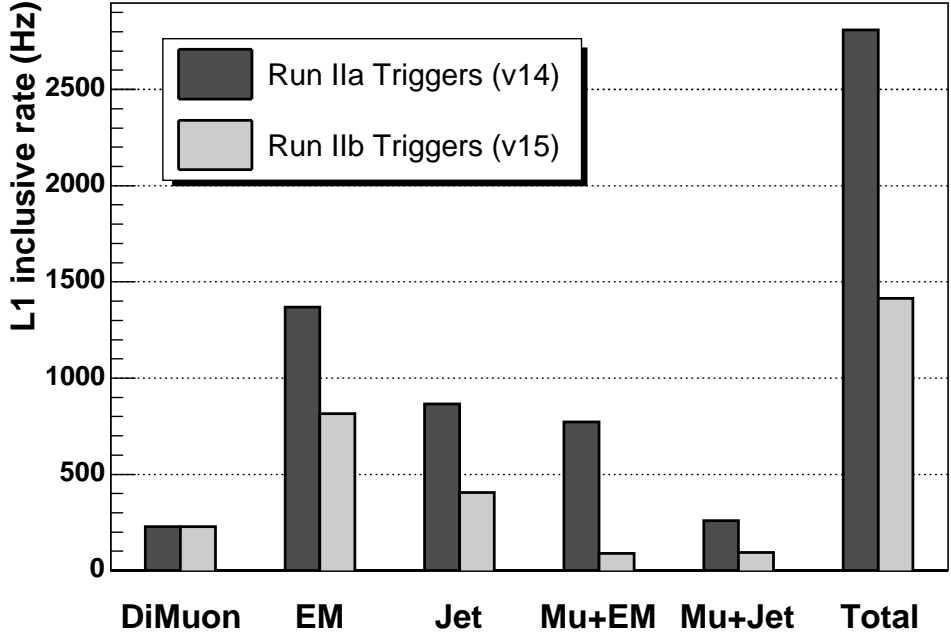}
  \caption{Predicted rates for Run IIa (v14) and Run IIb (v15) trigger
           lists, extrapolated to a luminosity of \scinot{2}{32}
           \instLunit\ from trigger-unbiased data collected at lower
           luminosity.} 
  \label{fig:rates}
\end{center}
\end{figure}

Figure \ref{fig:rates} shows the predicted rates at a luminosity
of \scinot{2}{32} \instLunit ,
estimated using the Trigger Rate Tool,
for trigger lists based on 
Run IIa algorithms (v14) and their
Run IIb equivalents (v15).
Both trigger lists were designed to give similar efficiencies for
physics objects of interest in Run IIb.
However, the Run IIb trigger list yields a rate approximately a factor
of two smaller than that achievable using Run IIa algorithms.

%==================================================================
% HARDWARE OVERVIEW
%==================================================================
\section{Hardware Overview}
\label{sect:hardware}

\begin{figure}
\begin{center}
  \mygraphics{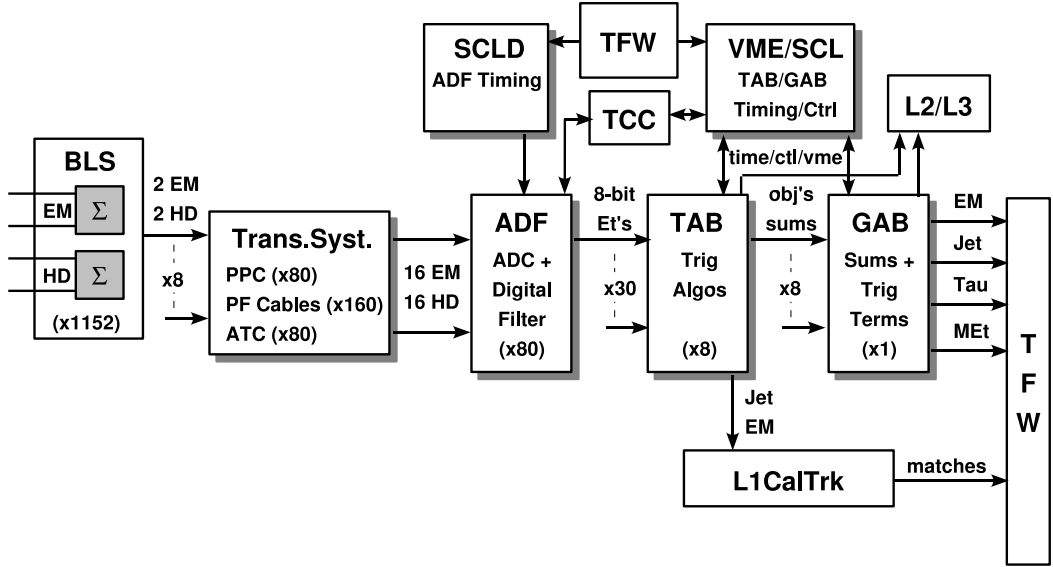}
  \caption{A block diagram of the main hardware elements of the Run
           IIb L1Cal system and their interconnections.}
  \label{fig:l1cal_blk}
\end{center}
\end{figure}

The algorithms described previously are implemented in several custom
electronics boards designed for the new L1Cal.
An overview of the main hardware elements of the Run IIb L1Cal system
is given in Fig.\ \ref{fig:l1cal_blk}.
Broadly, these elements are divided into three groups.
\begin{enumerate}
\item The {\it ADF System}, 
      containing those elements that receive and digitize analog
      TT signals from the BLS cards, and perform
      TT-based signal processing.
\item The {\it TAB/GAB System}, where algorithms are run on the
      digitized TT signals to produce trigger terms.
\item The {\it Readout System}, which inserts L1Cal information into
      the \Dzero\ data path for permanent storage.
\end{enumerate}

The L1Cal also communicates with other elements of the \Dzero\ trigger
and data acquisition (DAQ) system, including the following.
\begin{itemize}
\item The {\it Trigger Framework} (TFW), 
      which delivers trigger decisions and synchronizes the entire
      \Dzero\ DAQ.
      From the L1Cal point of view, the TFW
      sends global timing and control signals 
      (see Table \ref{table:tim-ctl})
      to the system over {\it Serial Command Links} (SCL) and
      receives the L1Cal and/or terms.
\item The L1Cal {\it Trigger Control Computer} (L1Cal TCC), 
      which configures and monitors the system.
\item The {\it Level-1 Cal-Track Match} trigger system (L1CalTrk),
      another L1 trigger system that performs azimuthal matching
      between L1CTT tracks and L1Cal EM and Jet objects.
\end{itemize}

\begin{table}
\begin{center}
  \caption{Timing and control signals used in the L1Cal
           system. Included are \Dzero\ global timing and control
           signals (SCL) used by the ADFs and the TAB/GAB system,
	   as well as intra-system communication and
           synchronization flags described later in the text.}
  \label{table:tim-ctl}
\vspace{1em}
\begin{tabular}{lccp{2.5in}}
\hline
  {\bf SCL} & {\bf ADF} & {\bf TAB/GAB}
  & {\bf Description} \\
\hline
  \init      & --- & yes & initialize the system \\
  \clk       & yes & yes & 132 ns Tevatron RF clock \\
  \firstturn & yes & yes & marks the first crossing of an accelerator
                           turn \\ 
  \realbx    & yes & --- & flags clock periods containing real beam
                           crossings \\
  \bxno      & --- & yes & counts the 159 bunch crossings in a turn \\
  \accept    & yes & yes & indicates that an L1 Accept has been issued
                           by the TFW \\
  \monitor   & yes & --- & initiates collection of ADF monitoring data\\
  \error     & --- & yes & a TAB/GAB error condition transmitted to
                           the SCL hub \\
  \busy      & --- & yes & asserted by the TABs/GAB until an observed
                           error is cleared \\
\hline
  ---        & \adfmon  & --- & allows TCC to freeze ADF circular
                                buffers \\
  ---        & \adftrig & --- & allows TCC to fake a \monitor\ signal
                                on the next L1 Accept \\
\hline
  ---        & --- & \tabrun  & TAB/GAB data path synchronization signal \\
  ---        & --- & \tabtrig & pulse to force writing to TAB/GAB
                                diagnostic memories \\
  ---        & --- & \tabfrm  & used for synchronization of TAB/GAB
                                VME data under VME/SCL control \\
  ---        & --- & \tabaddr & internal address for TAB/GAB VME
                                read/write operations \\
  ---        & --- & \tabdata & data for TAB/GAB VME read/write
                                operations \\ 
\hline
\end{tabular}
\end{center}
\end{table}

Within the L1Cal,
the ADF system consists of
the {\it Transition System},
the {\it Analog and Digital Filter} cards (ADF), 
and the {\it Serial Command Link Distributor} (SCLD). 
The Transition System,
consisting of {\it Patch Panels}, {\it Patch Panel Cards} (PPC),
{\it ADF Transition Cards} (ATC), and connecting cables,
adapts the incoming BLS signal cables to the
higher density required by the ADFs.
These ADF cards, 
which reside in four 6U VME-64x crates \cite{wiener},
filter, digitize and process individual TT signals, forming the building
blocks of all further algorithms. 
They receive timing and
control signals from the SCL via a
{\it Serial Command Link Distributor} card (SCLD).

Trigger algorithms are implemented in the L1Cal in two sets of cards:
the {\it Trigger Algorithm Boards} (TAB) and the
{\it Global Algorithm Board} (GAB),
which are housed in a single 9U crate with a custom backplane.
The TABs identify EM, Jet and Tau objects
in specific regions of the calorimeter
using the algorithms 
described in Section \ref{sect:algos} and also calculate partial
global energy sums.
The GAB uses these objects and energy sums to calculate
and/or terms, which the TFW uses to make trigger decisions.
Finally, the {\it VME/SCL card},
located in the L1Cal Control Crate,
distributes timing and control signals
to the TABs and GAB and provides a communication path for their
readout.

\begin{figure}
\begin{minipage}[t]{0.475\textwidth}
  \begin{center}
    \mygraphics{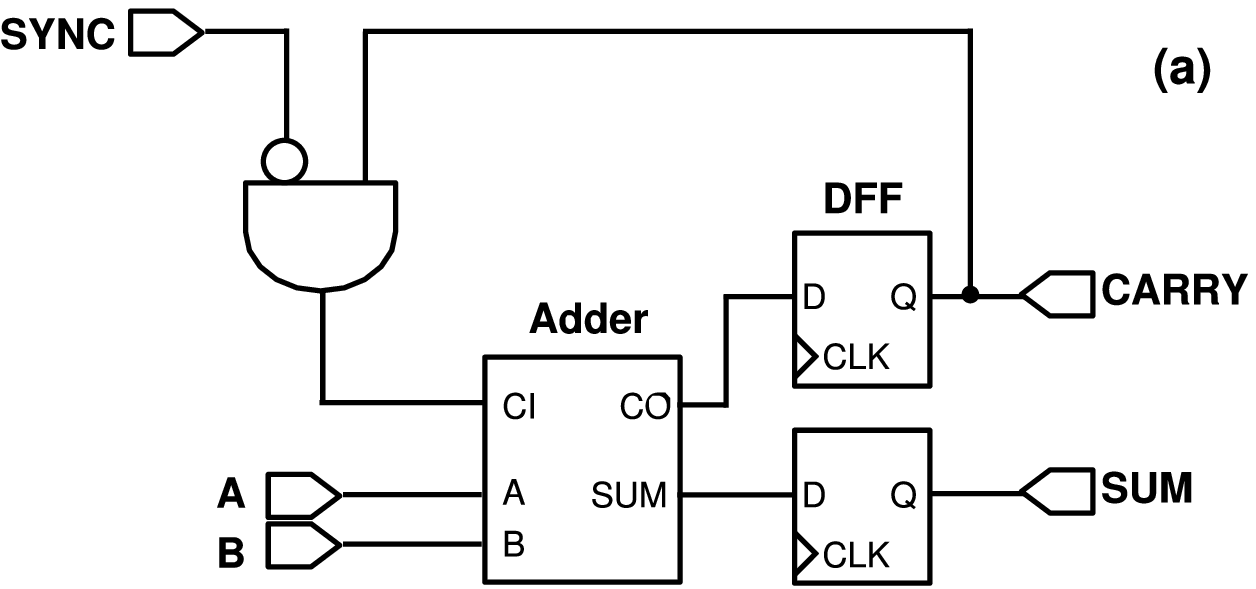}
  \end{center}
\end{minipage}
\hfill
\begin{minipage}[t]{0.475\textwidth}
  \begin{center}
    \mygraphics{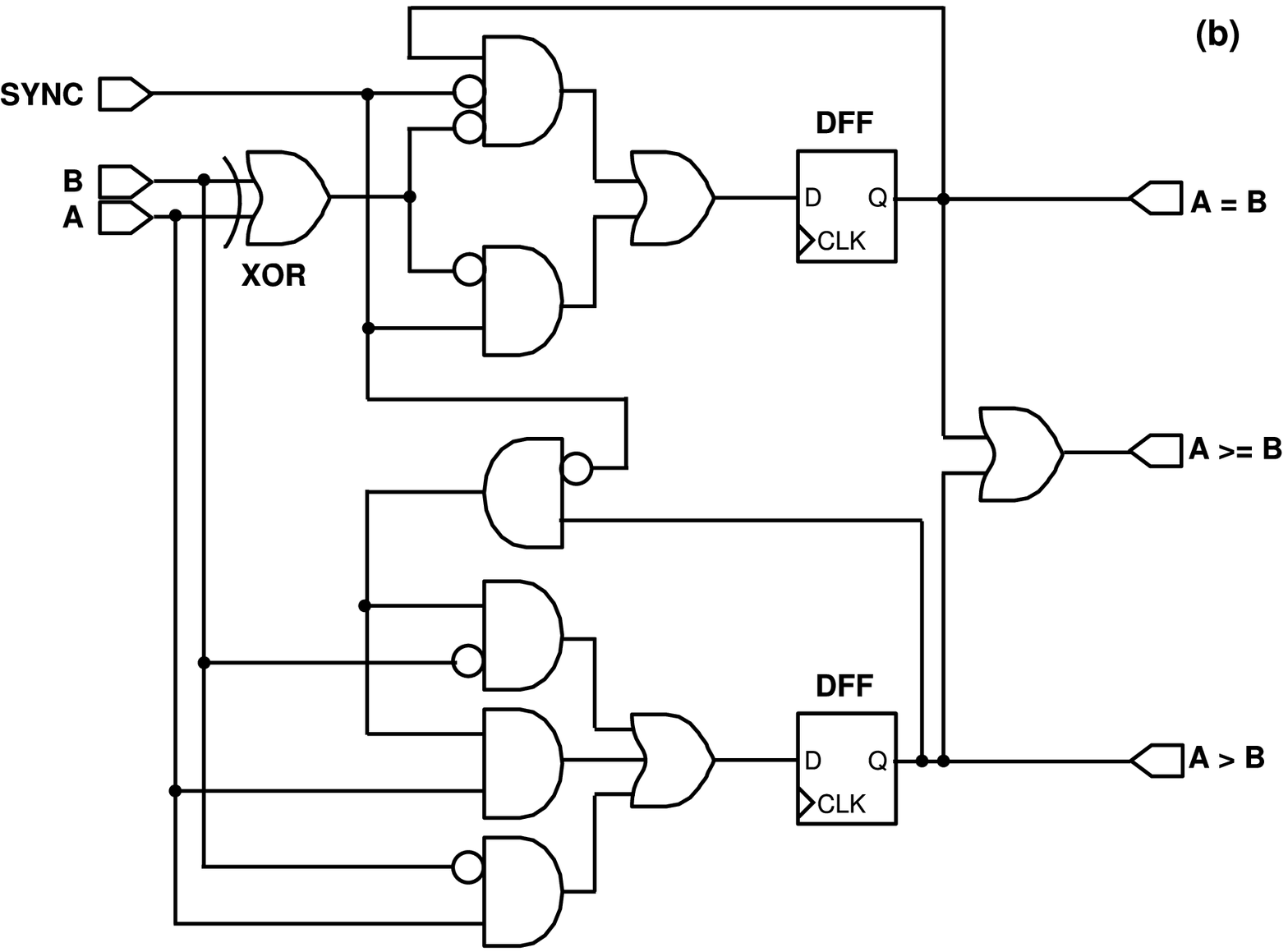}
  \end{center}
\end{minipage}
\begin{center}
  \caption{Logic diagrams for a Bit-serial adder (a) and
           a bit-serial comparator (b).}
  \label{fig:serial_arith} 
\end{center}
\end{figure}

The architecture of the L1Cal system
and the number of custom elements required,
summarized in Table \ref{table:boards},
is driven by the large amount of 
overlapping data required by the sliding windows algorithm.
In total, more than 700 Gbits of data per second are transmitted
within the system. Of this, each local maximum calculation requires
4.4 Gbits/s from 72 separate TTs. The most cost effective solution to
this problem, which still results in acceptable trigger decision
latency, is to deal with all data as serial bit-streams. Thus, all
intra-system data transmission is done bit-serially using the Low
Voltage Differential Signal (LVDS) protocol and nearly all algorithm
arithmetic is performed bit-serially as well, at clock speeds such
that all bits of a data word are examined in the 132 ns
Tevatron bunch crossing interval.
Examples of a bit-serial adder and comparator are shown in Fig.\
\ref{fig:serial_arith}. 
The only exception to this bit-serial
arithmetic rule is in the calculation of Tau object isolation, which
requires a true divide operation (see Section \ref{sect:algos}) and
thus introduces an extra 132 ns of latency to the trigger term
calculation. Even with this extra latency, the L1Cal results arrive at
the TFW well within the global L1 decision time budget.

\begin{table}
\begin{center}
  \caption{A summary of the main custom electronics elements of the
           L1Cal system. For each board, the TT region
	   (in \etaxphi ) that the board receives as input and sends on
           as output is given as well as the total number of each
           board type required in the system.}
  \label{table:boards}
\vspace{1em}
\begin{tabular}{lccc}
\hline
  {\bf Board} & {\bf Input TT Region} & {\bf Output TT Region}
  & {\bf Total Number} \\
\hline
  PPC     & \XxY{4}{4}   & \XxY{4}{4}  & 80 \\
  ATC     & \XxY{4}{4}   & \XxY{4}{4}  & 80 \\
\hline
  ADF     & \XxY{4}{4}   & \XxY{4}{4}  & 80 \\
  SCLD    & all          & all         & 1 \\
\hline
  TAB     & \XxY{40}{12} & \XxY{31}{4} & 8 \\
  GAB     & all          & all         & 1 \\
  VME/SCL & all          & all         & 1 \\
\hline
\end{tabular}
\end{center}
\end{table}

%==================================================================
% THE ADF SYSTEM
%==================================================================
\section{The ADF System}
\label{sect:adfsyst}

%------------------------------------------------------------------
\subsection{Transition System}
\label{sect:trans}

Trigger pick-off signals from the BLS cards of the
\EMtt\ and \HADtt\ calorimeters are
transmitted to the L1Cal trigger system, located in the Movable
Counting House (MCH), through 40--50 m long coaxial ribbon cables.
Four adjacent coaxial cables in a ribbon 
carry the differential signals from
the \EMtt\ and \HADtt\ components of a single TT. 
Since there are 1280 BLS trigger cables 
distributed among ten racks of the original L1Cal trigger
electronics,
the L1Cal upgrade was constrained
to reuse these cables. 
However, because the ADF input signal density is much larger than that in
the old system (only four crates are used to house the ADFs as
opposed to 10 racks for the old system's electronics)
the cables could not be
plugged directly into the upgraded L1Cal trigger electronics;
a transition system was needed.

The transition system is composed of passive electronics cards and
cables that route signals from the BLS trigger cables to
the backplane of the ADF crates
(see Section \ref{sect:adf}).  
It was designed to allow the trigger cables to remain
within the same Run I/IIa rack locations. 
It consists of the following elements.

\begin{itemize}
\item  Patch Panels and Patch Panel Cards (PPC):
       A PPC
       receives the input signals from 16
       BLS trigger cables and transmits the output through a pair of
       Pleated Foil Cables. A PPC also contains four connectors which
       allow the monitoring of the signals.
       Eight PPCs are mounted two to a Patch Panel
       in each of the 10 racks originally
       used for Run I/IIa L1Cal electronics.
\item Pleated Foil Cables:
      Three meter long
      Pleated Foil Shielded Cables (PFC), made by the 3M corporation
      \cite{pfc}, are used to 
      transfer the analog TT output signals from the PPC to the ADF
      cards via the ADF Transition Card.  There are two PFCs for each
      PPC for a total of 160 cables.  
      The unbalanced characteristic impedance specification of
      the PFC is 72 \Ohm , which provides a good impedance match to
      the BLS trigger cables.
\item ADF Transition Card (ATC): 
      The ATCs are passive cards connected to the ADF crate
      backplane. These cards receive the analog TT signals from two
      PFCs and transmit them to the ADF card.  There are 80 ATCs that
      correspond to the 80 ADF cards.  Each ATC also transmits the three
      output LVDS cables of an ADF card to the TAB crate -- a total
      of 240 LVDS cables.
\end{itemize}

%------------------------------------------------------------------
\subsection{ADF Cards}
\label{sect:adf}

The {\it Analog and Digital Filter} cards (ADF)
are responsible for sending 
the best estimate of the transverse energy (\Et ) 
in the \EMtt\ and \HADtt\ sections of
each of the 1280 TTs
to the eight TAB cards
for each Tevatron beam crossing.  
The calculation of these \Et\ values by the 80 ADF cards 
is based upon the 2560 analog trigger signals that the ADF cards 
receive from the BLS cards, and upon the timing and control
signals that are distributed throughout the \Dzero\ 
data acquisition system by the
Serial Command Links (SCL).  
The ADF cards themselves are 6U $\times$ 160 mm, 12-layer boards
designed to connect to a VME64x backplane using P0, P1 and P2
connectors.  
The ADF system is set up and monitored, over VME, by a
Trigger Control Computer (TCC),
described in Section \ref{sect:online}.

%------------------------------------------------------------------
\subsection{Signal Processing in the ADFs}
\label{sect:adf-proc}

\begin{figure}
\begin{center}
  \mygraphics{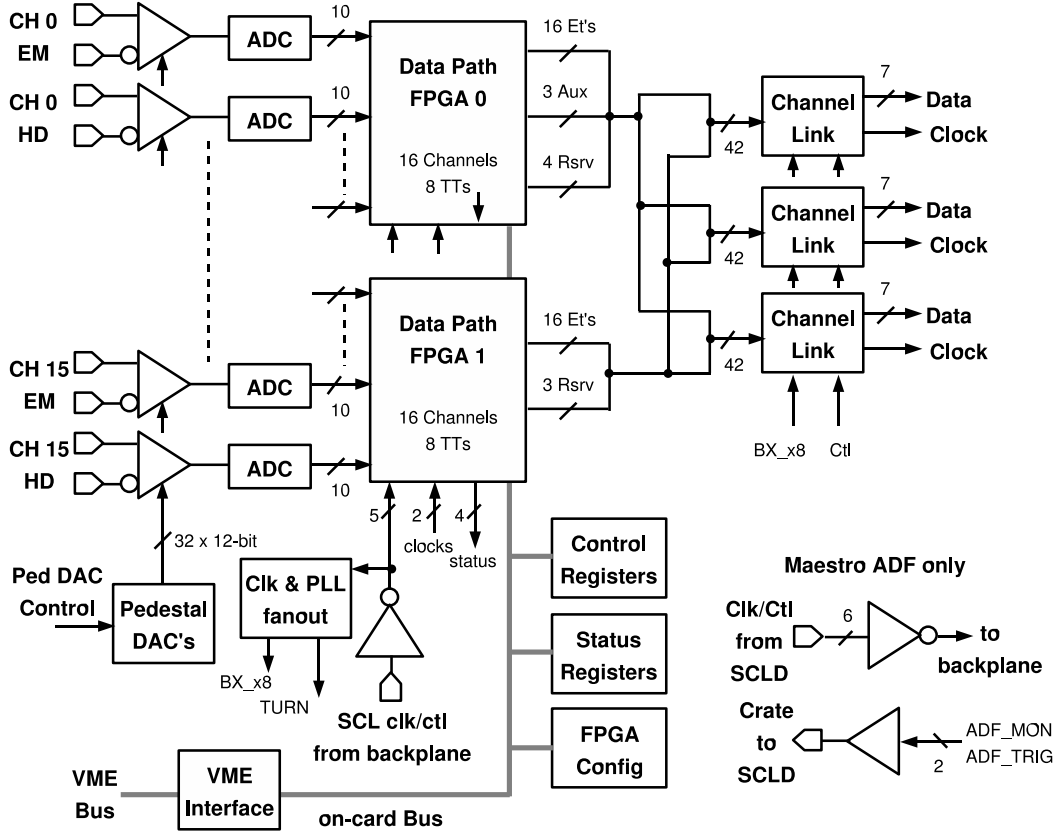}
  \caption{ADF card block diagram.}
  \label{fig:adf2_blk}
\end{center}
\end{figure}

Each ADF card,
as shown schematically in Fig.\ \ref{fig:adf2_blk},
uses 32 analog trigger signals corresponding to
the \EMtt\ and \HADtt\ components of a \XxY{4}{4}
array of Trigger Towers.  Each differential, AC coupled analog trigger
signal is received by a passive circuit that terminates and
compensates for some of the characteristics of the long cable that
brought the signal out of the collision hall.  Following this passive
circuit the active part of the analog receiver circuit rejects common
mode noise on the differential trigger signal, provides filtering to
select the frequency range of the signal caused by a real Tevatron
energy deposit in the Calorimeter, and provides additional scaling
and a level shift to match the subsequent ADC circuit.

The analog level shift in the trigger signal receiver circuit is
controlled, separately for each of the 32 channels on an ADF card, by
a 12 bit pedestal control DAC,
which can swing the output of the ADC that follows it from slightly
below zero to approximately the middle of its full range.
This DAC is used both to set the
pedestal of the signal coming out of the ADC that follows the receiver
circuit and as an independent way to test the full
signal path on the ADF card.
During normal operation, we set the pedestal at the ADC output to 50
counts which is a little less than 5\% of its full scale range.
This offset allows us to accommodate negative fluctuations in the
response of the BLS circuit to a zero-energy signal.

The 10 bit sampling ADCs \cite{adf-adc} 
that follow the receiver circuit 
make conversions every 33 ns --
four times faster than the Tevatron BX period of 132 ns.
This conversion rate is used 
to reduce the latency going through the pipeline ADCs and to provide
the raw data necessary to associate the rather slow rise-time trigger
signals (250 ns typical rise-time) with the correct Tevatron beam
crossing.  
Although associating energy deposits in the Calorimeter with the
correct beam crossing is not currently an issue since actual
proton-antiproton collisions only occur every 396 ns,
rather than every 132 ns as originally planned,
the oversampling feature has been retained for the flexibility it
provides in digital filtering.

\begin{figure}
\begin{center}
  \mygraphics{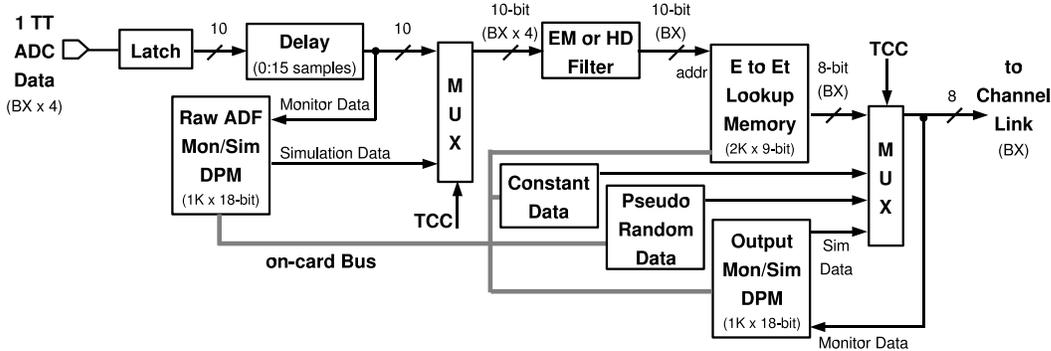}
  \caption{Block diagram of signal processing for a single TT in the
           ADF.} 
  \label{fig:adf2_data_proc}
\end{center}
\end{figure}

On each ADF card the 10 bit outputs from the 32 ADCs flow into a pair
of FPGAs \cite{adf-fpga}, 
called the {\it Data Path FPGAs}, where the bulk of the signal
processing takes place.
This signal processing task,
shown schematically in Fig.\ \ref{fig:adf2_data_proc},
is split over two FPGAs with each
FPGA handling all of the steps in the signal processing for 16
channels.  Two FPGAs were used because it simplified the circuit board
layout and provided an economical way to obtain the required number of
I/O pins.

The first step in the signal processing is to align in time all of the
2560 trigger signals.  The peak of the trigger signals from a given
beam crossing arrive at the L1Cal at different times
because of different cable lengths and different channel capacitances.
These signals are made isochronous using variable length shift
registers that can be set individually for each channel by the TCC.
Once the trigger signals have been aligned in time, they are sent to
both the Raw ADC Data Circular Buffers where monitoring data are
recorded and to the input of the Digital Filter stage.  

The Raw ADC Data Circular Buffers are typically set up to record 
all 636 of the ADC
samples registered in a full turn of the accelerator. 
This writing operation can be stopped by 
a signal from the TCC,
when an L1 Accept flagged with a special {\it Collect Status} flag is
received by the system on the SCL, or
in a {\it self-trigger} mode where any TT above a programmable
threshold causes writing of all Circular Buffers to stop.
Once writing has stopped, all data in the buffers can be read out
using the TCC, providing valuable monitoring information on the
system's input signals.
The Raw ADC Data Circular Buffers can also be loaded by
the TCC with simulated data,
which can be inserted into the ADF data path instead of real signals
for testing purposes.

The Digital Filter in the signal processing path can be used to remove
high frequency noise from the trigger signals and to remove low
frequency shifts in the baseline.  This filter is currently configured
to select the ADC sample at the peak of each analog TT signal.
This mode of operation allows the most direct comparison with data
taken with the previous L1Cal and appears to be adequate for the
physics goals of the experiment.

The 10 bit output from the Digital Filter stage 
has the same scale and offset as the output from the ADCs.  
It is used as an address to an {\it E to \Et\ Lookup Memory},
the output of which is an eight bit data word
corresponding to the \Et\ seen in that TT.
This $E$ to \Et\ conversion is normally programmed such that one output
count corresponds to 0.25 GeV of \Et\ and includes an eight count
pedestal, corresponding to zero \Et\ from that TT.

The eight bit TT \Et\ is one of four sources of data that can be sent
from the ADF to the TABs under control of a multiplexer
(on a channel by channel and cycle by cycle basis). 
The other three multiplexer inputs are 
a fixed eight-bit value read from a programmable register,
simulation data from the Output Data Circular Buffer,
and data from a pseudo-random number generator.

The latter two of these sources are used for system testing
purposes. During normal operation, the multiplexers are set up such
that TT \Et\ data are sent to the TABs on those bunch crossing
corresponding to real proton-antiproton collisions, while the fixed
pedestal value (eight counts) is sent on all other accelerator clock
periods. If noise on a channel reaches a level where it significantly
impacts the \Dzero\ trigger rate, then this channel can be disabled,
until the problem can be resolved,
by
forcing it to send the fixed pedestal on all accelerator clock
periods, regardless of whether they contain a real crossing or
not. Typically, less than 10 (of 2560) TTs are excluded in this
manner at any time.

Data are sent from the ADF system to the TAB cards using a National
Semiconductor Channel Link chip set with LVDS signal levels between
the transmitter and receiver \cite{chan-link}.
Each Channel Link output from an ADF
card carries the \Et\ data for all 32 channels
serviced by that card.  A new frame of \Et\ data is sent every 132 ns.
All 80 ADF cards begin sending their frame of data for a given
Tevatron beam crossing at the same point in time.  
Each ADF card sends out three identical copies of its data to
three separate TABs, accommodating the data sharing requirements of the
sliding windows algorithm.

%------------------------------------------------------------------
\subsection{Timing and Control in the ADF System}
\label{sect:adf-time}
The ADF system receives timing and control
signals listed in Table \ref{table:tim-ctl}
over one of the SCLs \cite{d0run2}.
Distribution of these signals from the SCL to the 80 ADF cards is
accomplished by the {\it SCL Distributor} (SCLD) card.
The SCLD card receives a copy of the SCL information using a 
\Dzero -standard
SCL Receiver mezzanine card
and fans out the signals mentioned in Table \ref{table:tim-ctl} 
to the four VME-64x crates that hold the 80 ADF cards using LVDS level
signals. 
In addition, each ADF crate
sends two LVDS level signals
(\adfmon\ and \adftrig )  back to the SCLD card,
allowing the TCC to cause synchronous readout of the ADFs.

Within an ADF crate, the ADF card at the mid-point of the backplane
(referred to as the {\it Maestro})
receives the SCLD signals and places
them onto spare, bused VME-64x backplane lines at TTL open
collector signal levels. All 20 of the ADF cards in a crate pick up 
their timing and control signals from these backplane lines. To ensure
a clean clock, the \clk\ signal is sent differentially
across the backplane and is used as the reference for a PLL on the
ADFs. This PLL provides the jitter-free clock signal needed for LVDS
data transmission to the TABs and for ADC sampling timing.

%------------------------------------------------------------------
\subsection{Configuring and Programming the ADF System}
\label{sect:adf-prog}
The ADF cards are controlled over a VME bus using a VME-slave
interface implemented in a PAL that is automatically configured at
power-up. 
Once the VME interface is running, 
the TCC simultaneously loads identical logic files into the two data path
FPGAs on each card.
Since each data path FPGA uses slightly different logic
(\eg , the output check sum generation),
the FPGA {\it flavor} is chosen by a single ID pin.
After TCC has configured all of the data path FPGAs,
it then programs all
control-status registers and memory blocks in the ADFs.
Information
that is held on the ADF cards that is critical to their physics
triggering operation is protected by making those programmable
features ``read only'' during normal operation.  TCC must explicitly
unlock the write access to these features to change their control
values.  In this way no single failed or mis-addressed VME cycle can
overwrite these critical data.

%==================================================================
% ADF-TO-TAB DATA TRANSFER
%==================================================================
\section{ADF to TAB Data Transfer}
\label{sect:adf-to-tab}

\begin{figure}
\begin{center}
  \mygraphics{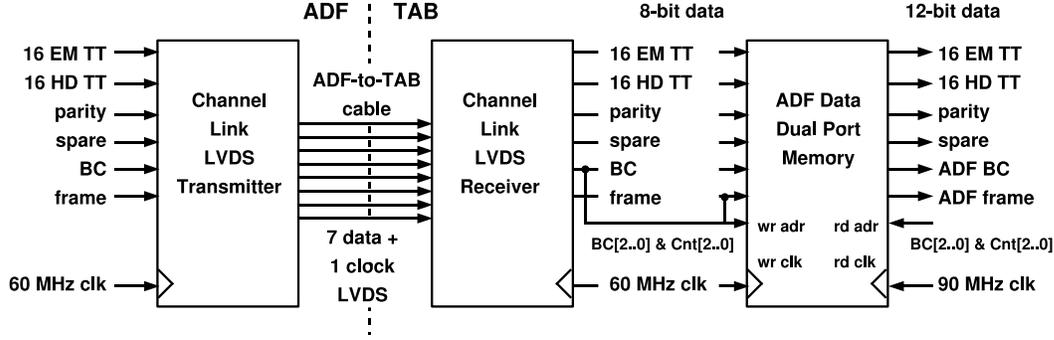}
  \caption{ADF to TAB data transmission, reception and the
           dual-port-memory transition from 8-bit to 12-bit data.}
  \label{fig:adf_to_tab}
\end{center}
\end{figure}

Digitized TT data from each ADF's \XxY{4}{4}, \etaxphi\
region are sent to the TABs for further processing, 
as shown in Fig.\ \ref{fig:adf_to_tab}.
To accommodate the
high density of input on the TABs, the 8-bit serial trigger-tower data are 
transmitted using the 
channel-link LVDS chipset \cite{chan-link},
 which serializes 48 CMOS/TTL inputs and the
transmission clock onto seven LVDS channels plus a clock channel.  
In the L1Cal system,
the input to the transmitter is 60 MHz TTL 
(eight times the bunch crossing rate), 
which is stepped up to 420 MHz for LVDS transmission.

Each ADF sends three identical copies of 36 8-bit words to three
different TABs on each bunch crossing. 
This data transmission 
uses eight LVDS channels --
seven data channels containing six serialized data words each, and one
clock --
on Gore cables with 2mm HM connectors \cite{gore}.
The 36 data words consist of
the digitized \Et\ of 16 \EMtt\ and 16 \HADtt\ TTs
and four control words. 
The bunch crossing number
control word indicates which accelerator crossing produced the ADF
data being transmitted,
and is used throughout the system for synchronization.  
The frame-bit
control word is used to help align the least significant bits of the other
data words.  
The parity control word is the logical XOR of every other word
and is used to check the integrity of the data transmission. 
Finally, one control word is reserved for future use.

While the ADF logic is 8-bit serial (60 MHz) the TAB logic is 12-bit serial
(90 MHz).  To cross the clock domains, the data passes through a dual-port
memory with the upper four bits padded with zeros.
The additional bit space is
required to accommodate the sliding windows algorithm sums.

The dual port memory write address is calculated from 
the frame and bunch crossing words of
the ADF data.  The least significant address bits are a data word bit
count, which is reset by the frame signal, 
while the most significant address bits are the first three-bits of the bunch
crossing number.  This means that the memory is large enough to contain
eight events of eight-bit serial data.

By calculating the read address in the same fashion, but from the TAB frame
and bunch crossing words, 
the dual-port memory crosses 60~MHz/90~MHz
clock domains, maintains the correct phase of the data, and synchronizes the
data to within eight crossings all at the same time.  This means the TAB timing
can range between a minimal latency setting where the data are retrieved just
after they are written
and a maximal latency setting where the data are
retrieved just before they are overwritten.
If the TAB timing is outside this
range, the data from eight previous or following crossings 
will be retrieved.

Although off-the-shelf components were used within their specifications, 
operating 240 such links reliably was found to be
challenging. Several techniques were employed to stabilize the data
transmission. Different cable lengths 
(between 2.5 and 5.0 m)
were used to match the different distances between ADF crates and the
TAB/GAB crate. 
The DC-balance and pre-emphasis features
of the channel-link chipset \cite{chan-link}
were also used, 
but deskewing, which was found to be unreliable, was not.

%==================================================================
% THE TAB/GAB SYSTEM
%==================================================================
\section{The TAB/GAB System}
\label{sect:tab-gab}

%------------------------------------------------------------------
\subsection{Trigger Algorithm Board}
\label{sect:tab}

\begin{figure}
\begin{center}
  \mygraphics{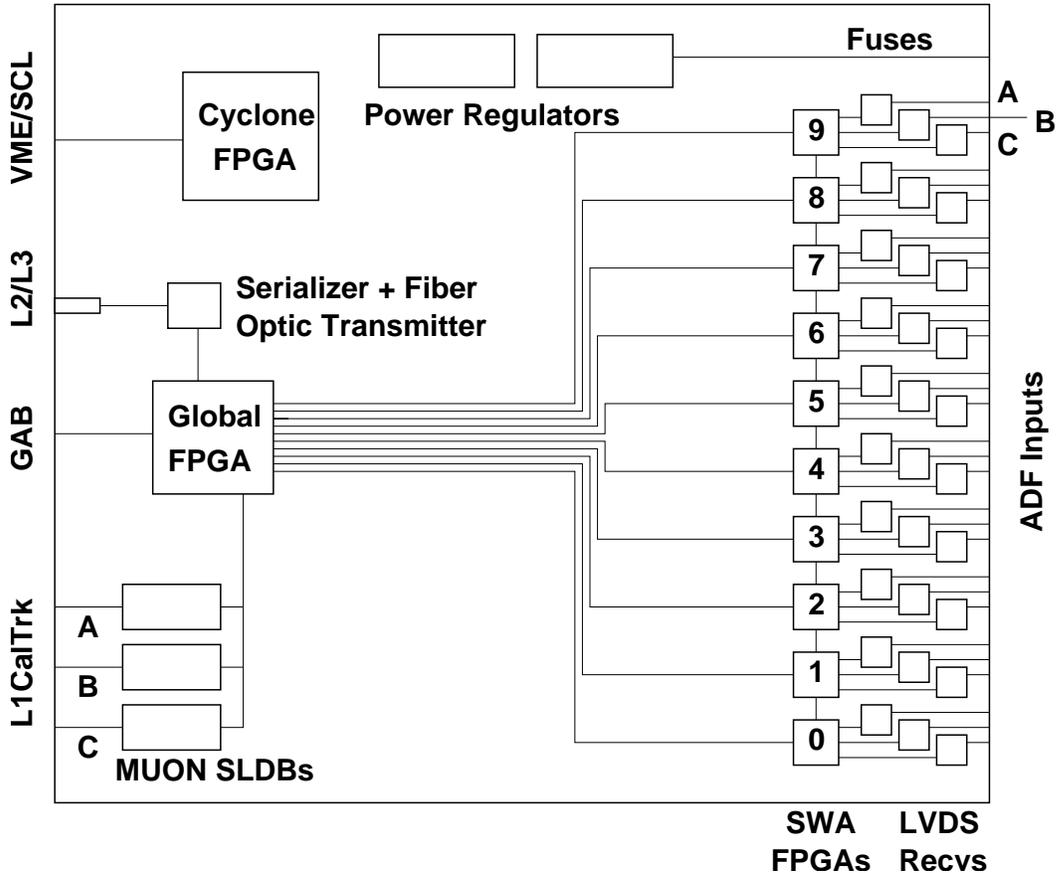}
  \caption{Block diagram of the TAB.}
  \label{fig:tab_overview}
\end{center}
\end{figure}

The Trigger Algorithm Boards (TABs) find EM, Jet and Tau candidates
using the sliding windows algorithm and perform preliminary sums for
total and missing \Et\ calculations.
Each TAB is a double-wide 9U $\times$ 400 mm, 12-layer
card designed for a custom backplane.
The main functional elements of the TAB are shown in  Fig.\
\ref{fig:tab_overview}. 

In the TAB's main trigger data path,
LVDS cables from 30 ADFs are received at the back of the card using
feedthrough connectors on the backplane. 
The data from these cables are extracted using Channel Link
receivers\cite{chan-link}  
and sent,
as individual bit-streams for each TT,
to ten TAB sliding windows
algorithm (SWA) FPGAs \cite{stratix} for processing.
These chips also pass some of their data to their nearest neighbors to
accommodate the data sharing requirements of the sliding windows
algorithms. 
The algorithm output from each SWA is sent to a single
TAB global FPGA \cite{stratix}.  
The global FPGA calculates regional sums and sends the
results out the front of the board to the GAB, 
over the same type of cables used for ADF to TAB data transmission
(see Section \ref{sect:adf-to-tab})
using embedded LVDS functionality in the FPGA.
This data transmission occurs at a clock rate of 636 MHz.

The global FPGA also sends three copies of Jet and EM object
information for each bunch crossing
to the L1CalTrk system for processing using 
Gbit/s serial link transmitter daughter cards (MUON SLDB) \cite{d0run2}.  
Upon receiving an L1 accept from the \Dzero\ TFW,
the TAB global chip also sends data out
on a serial fiber-optic link \cite{optic}
for use by the L2 trigger 
and for inclusion in the \Dzero\ event data written to permanent
storage on an L3 accept.

Low-level board services are provided by the TAB Cyclone chip
\cite{cyclone},
which is configured by an on-board serial configuration device
\cite{cyc-conf} on TAB power-up. 
These services include providing the path for power-up and
configuration of the other FPGAs on the board, under the direction of
the VME/SCL card; communicating with VME and the \Dzero\ SCL 
over the specialized VME/SCL serial link;
and fanning out the 132 ns detector clock using an on-board clock
distribution device \cite{clk-dist}.

%------------------------------------------------------------------
\subsection{Global Algorithm Board}
\label{sect:gab}

\begin{figure}
\begin{center}
  \mygraphics{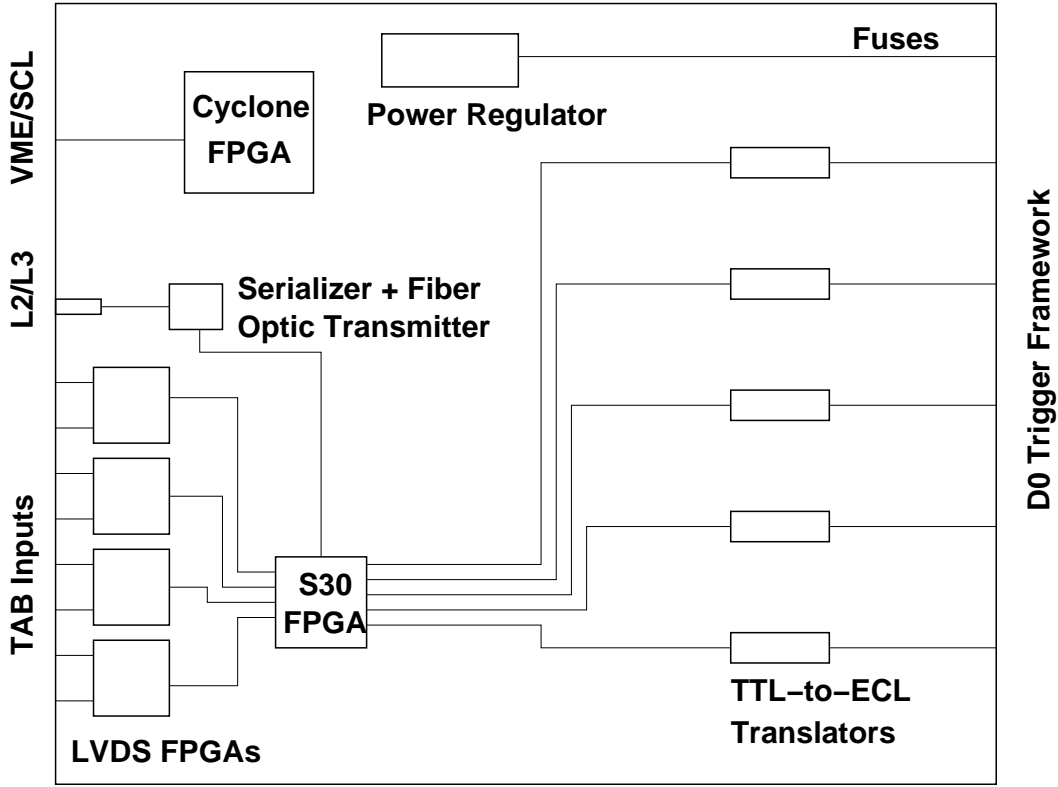}
  \caption{Block diagram of the GAB.}
  \label{fig:gab_overview}
\end{center}
\end{figure}

The global algorithm board (GAB)
receives data containing regional counts of Jet, EM, and Tau physics
objects calculated 
by the TABs and produces a menu of and/or terms,
which is sent to the \Dzero\ trigger framework.  
Like the TAB, the GAB
is a double-wide 9U $\times$ 400 mm, 12-layer circuit board 
designed for a custom backplane. 
Its main functional elements are shown in Fig.\ \ref{fig:gab_overview}.

LVDS receivers,
embedded in four Altera Stratix FPGAs (LVDS FPGAs) \cite{stratix} 
each receive the output of two TABs, synchronizing the
data to the GAB 90 MHz clock using a dual-port memory.  
The synchronized TAB data from all four LVDS FPGAs is sent to a
single GAB S30 FPGA \cite{stratix},
which calculates and/or terms, and sends them to the
trigger framework through TTL-to-ECL converters \cite{ttl-ecl}.  
There are five 16-bit
outputs on the GAB, although only four are used by the framework.

Much like the TABs, upon receiving an L1 accept, 
the GAB S30 sends data to L2 and L3
on a serialized fiber-optic link \cite{optic}.  
Also as on the
TABs, a Cyclone FPGA \cite{cyclone} provides low-level board services.

%------------------------------------------------------------------
\subsection{VME/SCL Board and the TAB/GAB Control Path}
\label{sect:vmescl}

\begin{figure}
\begin{center}
  \mygraphics{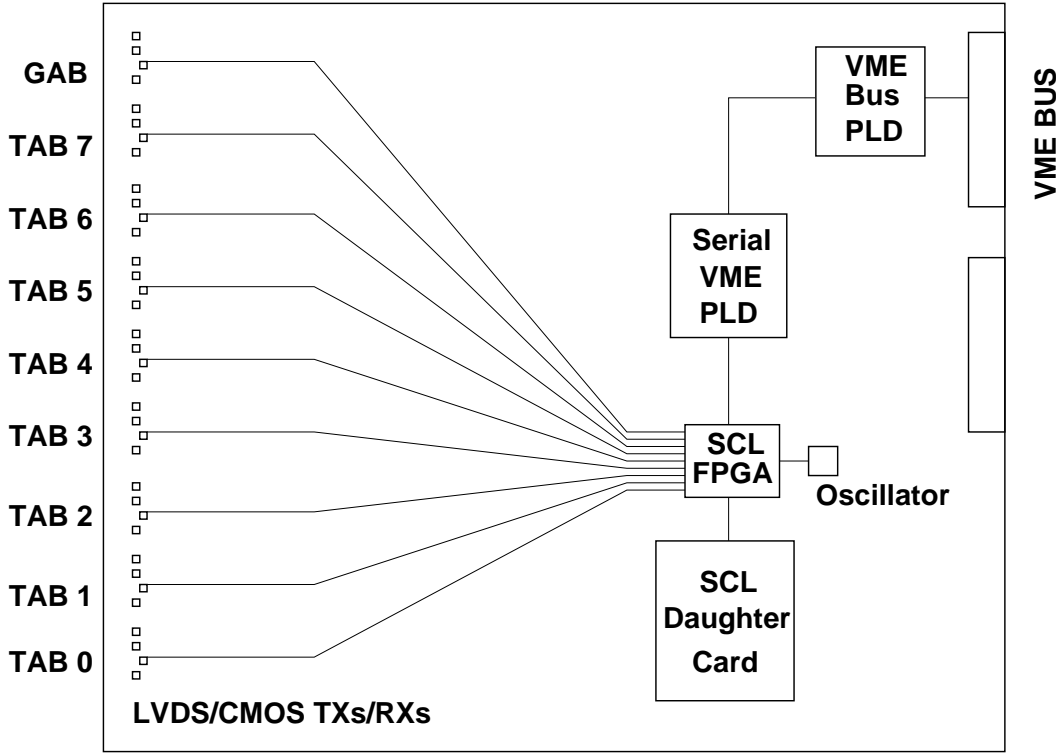}
  \caption{Block diagram of the VME/SCL board.}
  \label{fig:vmescl_overview}
\end{center}
\end{figure}

Because of the high-density of inputs to the TAB and GAB modules, 
direct connections of these cards to a VME bus is impossible.
A custom control path for these boards is
provided by the VME/SCL module,
a double-wide 9U $\times$ 400 mm, 8-layer board.
A block diagram of the main elements of this card can be found in 
Fig.\ \ref{fig:vmescl_overview}.
SCL signals arrive at the VME/SCL board via an
SCL receiver daughter card and those signals used by the
TAB/GAB system are selected for fanout by the SCL FPGA \cite{ep1k50}, 
which also
handles transmission/reception of serialized VME communications with
the TABs and GAB.
Any VME communication, directed to (from) a card in the TAB/GAB system, 
is received by (transmitted from) the VME Bus PLD,
which implements the VME protocol.
Those commands whose destination (source) is one of the TAB or GAB
boards are translated to (from) the custom serial protocol listed in
Table \ref{table:tim-ctl} by the Serial VME PLD, which connects to the
SCL FPGA for signal transmission (reception).
Serial communications between the VME/SCL card and the TABs and GAB is
accomplished using LVDS protocol \cite{vme-lvds}, on nine cables --
one for each TAB and GAB.

%------------------------------------------------------------------
\subsection{TAB/GAB Trigger Data Path}
\label{sect:tabgab-data}

\begin{figure}
\begin{center}
  \mygraphics{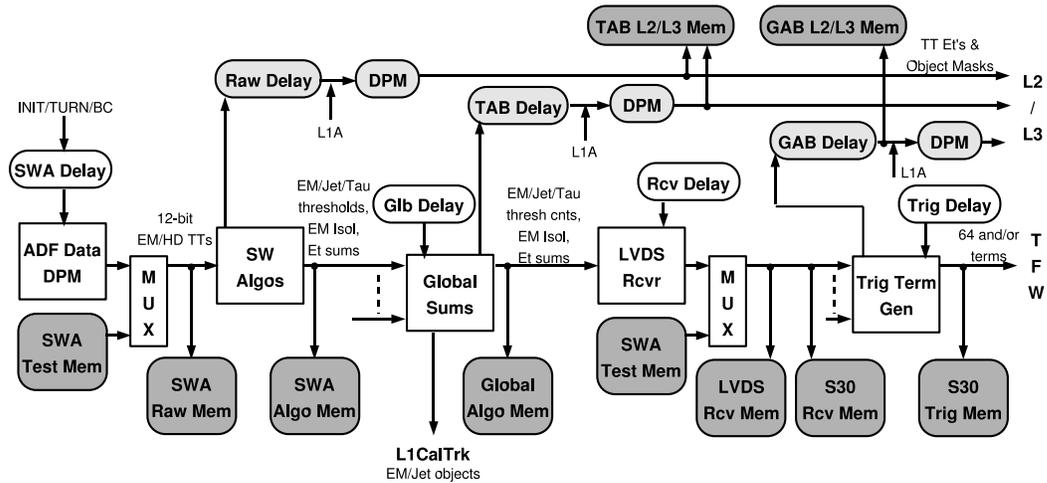}
  \caption{Data and Timing in the TAB/GAB system. 
    The {\it trigger path} consists of those elements with white
    background, while the {\it readout path} is highlighted in light
    gray and the {\it test memories} have darker gray backgrounds.}
  \label{fig:tab_rdout}
\end{center}
\end{figure}

The path of {\it trigger data} through the TAB/GAB system is shown in
Fig.\ \ref{fig:tab_rdout}. 
Each of the eight TABs receives data from 30 ADF cards,
covering a \XxY{40}{12} region in \etaxphi\ space.
Eight-bit TT \Et\ data are translated to 12-bit words in the ADF Data
DPM and are transmitted serially to the SWA FPGAs where EM, Jet and Tau
objects are  found.
Each of the ten TAB SWA chips finds objects in a \XxY{4}{4},
\etaxphi\ grid,
for which it requires a \XxY{9}{9} region of input TTs.
This TT data comes from the three LVDS receivers (A,B,C in Fig.\
\ref{fig:tab_overview}) attached directly to
the chip and also, indirectly, from its nearest neighbor SWA chips.
A map of
the TT inputs to a single SWA chip is given in Table
\ref{table:tt_index}.
In this table and the following discussion, 
we use {\it global indices}
($\eta[0,39]$ and $\phi[0,31]$) 
when referring to the entire grid 
but switch to {\it local indices} 
(\deta [$-$2,6] and \dphi [$-$2,6]) 
for single SWA chips.
The translation between the two systems is given below.
\begin{eqnarray}
\phi &=& 4 \times ({\rm TAB~No.}) + \dphi , \nonumber \\
\eta &=& 4 \times ({\rm SWA~chip~No.}) + \deta . 
\end{eqnarray}
Note that data for $\eta$ indices 0, 1, 38, and 39, at all $\phi$
positions, correspond to signals from the ICR detectors, which can be
added to the relevant calorimeter TTs if desired.

\begin{table}
\begin{center}
  \caption{TT input to a single TAB SWA chip. The TT grid is labeled
           in the SWA chip local coordinates, 
	   \deta\ (row),\dphi\ (column), while
           individual TTs are labeled, 0--81, as they are used in the
           firmware.}
  \label{table:tt_index}
\vspace{1em}
\begin{tabular}{|r|rr|rrrr|rrr|c|}
\hline
  cable:            &  \multicolumn{2}{|c|}{\it C} 
                    &  \multicolumn{4}{|c|}{\it B} 
                    &  \multicolumn{3}{|c|}{\it A}
                    & chip \\
\hline
  \deta /\dphi      & $-${\it 2} & $-${\it 1} &  {\it 0} &  {\it 1} 
                    & {\it 2} &  {\it 3} &  {\it 4} &  {\it 5} &  {\it 6} & \\
\hline
  {\it 6}           & 72 & 73 & 74 & 75 & 76 & 77 & 78 & 79 & 80 & \\ 
  {\it 5}           & 63 & 64 & 65 & 66 & 67 & 68 & 69 & 70 & 71 
                    & SWA $i+$1 \\ 
  {\it 4}           & 54 & 55 & 56 & 57 & 58 & 59 & 60 & 61 & 62 & \\ 
\hline
  {\it 3}           & 45 & 46 & 47 & 48 & 49 & 50 & 51 & 52 & 53 & \\ 
  {\it 2}           & 36 & 37 & 38 & 39 & 40 & 41 & 42 & 43 & 44 
                    & SWA $i$ \\ 
  {\it 1}           & 27 & 28 & 29 & 30 & 31 & 32 & 33 & 34 & 35 & \\ 
  {\it 0}           & 18 & 19 & 20 & 21 & 22 & 23 & 24 & 25 & 26 & \\ 
\hline
  $-${\it 1}        &  9 & 10 & 11 & 12 & 13 & 14 & 15 & 16 & 17 
                    & SWA $i-$1 \\ 
  $-${\it 2}        &  0 &  1 &  2 &  3 &  4 &  5 &  6 &  7 &  8 & \\ 
\hline
\end{tabular}
\end{center}
\end{table}

Each SWA chip sends the results of its algorithms to the Global Chip as
12-bit serial data on 25 lines. The data transmitted consists of the
following. 
\begin{itemize}
\item The highest of seven possible \Et\ thresholds passed by EM and Jet
      objects at each of the \XxY{4}{4}, \etaxphi\ positions
      considered by this chip, or zero if the object \Et\ is below all
      thresholds. This information (three bits for each position and
      object) is packed into a total of eight, 12-bit words, with each
      word containing data from the four $\eta$ locations at a
      specific $\phi$ for one object type.
\item The highest of seven possible Tau isolation ratio thresholds 
      (see Section \ref{sect:taualgo}) passed by Tau
      objects at each of the \XxY{4}{4}, \etaxphi\ positions
      considered by this chip, or zero if the ratio is below all
      thresholds. This information is packed in the same way as the EM
      and Jet object data above.
\item The results of the EM isolation and EM fraction calculations
      (see Section \ref{sect:emalgo}) for each of the \XxY{4}{4}
      locations considered in this chip.
      A single bit, corresponding to a specific \deta ,\dphi\ location
      is set if the EM object at that location passes both the EM
      isolation and EM fraction cuts.
\item Sums over four $\eta$ locations of \EMtt +\HADtt\ \Et\ 
      for each $\phi$ position considered in this chip.
\item Four-bit counts of the number of TTs in the chip with \EMtt
      +\HADtt\ \Et\ greater than three programmable thresholds. This
      information is used to aid in the identification of noisy
      channels. 
\item Raw TT \Et 's for transmission to the L2 and L3 systems (only
      transmitted on those BCs marked as \accept ).
\item The bunch crossing number
      and a flag indicating if there was a bunch crossing
      number mismatch between the ADF data and the TAB's local BX .
\item Status information.
\item Two spare lines.
\end{itemize}

The Global Chip receives these data from the ten SWA Chips and
constructs object counts 
in a \XxY{31}{4} region,
as well as \Et\ sums.
The reduced number of positions available for TAB object output
comes from edge effects in the sliding windows algorithms and from the
use of TTs at $\eta$ indices 0, 1, 38, and 39 for ICR energies.
The TABs further concentrate their data by summing object counts in
three $\eta$ ranges -- 
North (N), Central (C), and South (S) \cite{regions} -- 
before sending their results to the GAB.

A total of 48 12-bit data words are transmitted from each TAB to the
GAB. These data include the following.
\begin{itemize}
\item Two-bit counts of the number of EM and Jet objects over
      each of six possible \Et\ thresholds in the N, C, and S regions
      for each of the four $\phi$ positions considered by the
      TAB. Each 12-bit word contains counts for all six thresholds for
      a specific object in an $\eta$ region and $\phi$ position.
\item Two-bit counts of the number of Tau objects over
      each of six possible Tau ratio thresholds in the same format as
      the EM and Jet information above.
\item Single bits indicating that at least one EM object passed the
      isolation criteria in an $\eta$ region (S,C,N) at a specific
      $\phi$ position. Since not enough data lines were available to
      transmit isolation information for each possible EM object, this
      grouping represents a compromise that allows the GAB to
      construct isolated EM triggers if {\it any} EM object in an $\eta$
      region is found to be isolated.
\item Sums of \EMtt +\HADtt\ \Ex , \Ey , and scalar \Et\ over
      the \XxY{40}{4} region belonging to the TAB. \Ex\ and \Ey\ are
      calculated using sine and cosine look-ups appropriate for each
      TT's $\phi$ position.
\item Eight-bit counts of the number of TTs with \EMtt +\HADtt\ \Et\
      greater than three thresholds.
\item Bunch Crossing, status, synchronization and parity information.
\end{itemize}

At the GAB, TAB data are received and transmitted unchanged to the S30
Chip where and/or trigger terms are constructed as described in
Section \ref{sect:trig-list}. A total of 64 and/or terms are sent from
the GAB to the Trigger Framework.

%------------------------------------------------------------------
\subsection{TAB/GAB Timing and Readout}
\label{sect:tabgab-readout}

The timing and readout of the TAB and GAB modules,
shown in Fig.\ \ref{fig:tab_rdout}, are interrelated.
Both data traveling on the {\it trigger path} and on the 
{\it readout path} to the L2 and L3 systems on \accept\ 
must be synchronized so
that they correspond to a single, known bunch crossing number.
This synchronization is accomplished by setting adjustable {\it Delay}
FIFOs in the TABs and GAB such that the \bxno\ stamp on the data at
each stage in processing corresponds to the \bxno\ being transmitted
to the TAB/GAB system by the VME/SCL card. Errors are 
stored in status registers if a
mismatch between these numbers is detected at any point in the chain.

Readout of TAB/GAB data for further processing in the L2 and L3
trigger systems is accomplished by storing data, 
at various stages of the processing,
in pipelines ({\it Raw Delay}, {\it TAB Delay}, and {\it GAB Delay}),
whose depth is adjusted so that the data appears at the end of the
pipeline when the L1 trigger decision arrives at the boards.
If the decision is \accept , then the relevant data are
moved to Dual Port Memory buffers for
transmission, via optical fiber, to the L2 and L3 systems.

Identical data are sent to L2 and L3 by optically splitting the output
signals. These data consist of the following:
\begin{itemize}
\item The raw eight-bit \EMtt\ and \EMtt +\HADtt\ \Et\ values for each
      TT ({\it Raw}). 
\item A bit-mask with each bit corresponding to a possible EM, Jet or
      Tau object either set or not depending on whether the object has
      passed a L2 \Et\ threshold ({\it TAB}).
\item The set of 64 and/or terms transmitted from the GAB and the
      total \Et , \Ex , and \Ey\ sums ({\it GAB}).
\item A set of control, status and data integrity checksum words.
\end{itemize}

%------------------------------------------------------------------
\subsection{TAB/GAB data to L1CalTrk}
\label{sect:caltrk-data}

The L1CalTrk system receives
EM and Jet object data for each $\phi$ position from the 
TAB Global Chips \cite{caltrkdata}.
Each TAB sends three identical copies of its data 
(to eliminate cracks in the acceptance)
to the L1CalTrk system using three 
Muon Serial Link Daughter cards.
These daughter cards serialize seven 16-bit words per bunch crossing
period and transmit them to Muon Serial Link Receiver Daughter cards
in the L1CalTrk electronics. Four of these words contain EM and Jet
information for each of the $\phi$ regions
considered by the TAB. Each word is broken into seven-bit EM and Jet
parts, where bit $i$ is set in each part if any object of that type
above threshold $i$ is found in the full $\eta$ range. A parity  word
and two spare words are also transmitted.

%------------------------------------------------------------------
\subsection{TAB/GAB diagnostic memories}
\label{sect:tabgab-diagnostic}

The TAB and GAB modules have a series of VME-readable diagnostic
memories (see Fig.\ \ref{fig:tab_rdout})
designed to capture data from each step of the algorithm calculation.
Their contents are snapshots of data transfered between elements of
the TAB/GAB system and are generally capable of holding data for 32
consecutive bunch crossing periods,
although the {\it L2/L3 Memories} and the {\it S30 Trig Memory} are
limited to one event's worth of data.
These memories are normally written when a \tabtrig\ signal is sent
from the VME/SCL board under user control.
Both the TABs and GAB also have VME-writable test input memories, which
allow arbitrary patterns to be used in the place of the incoming data
from the ADF or TAB cards.

%==================================================================
% ONLINE CONTROL
%==================================================================
\section{Online Control}
\label{sect:online}

\begin{figure}
\begin{center}
  \mygraphics{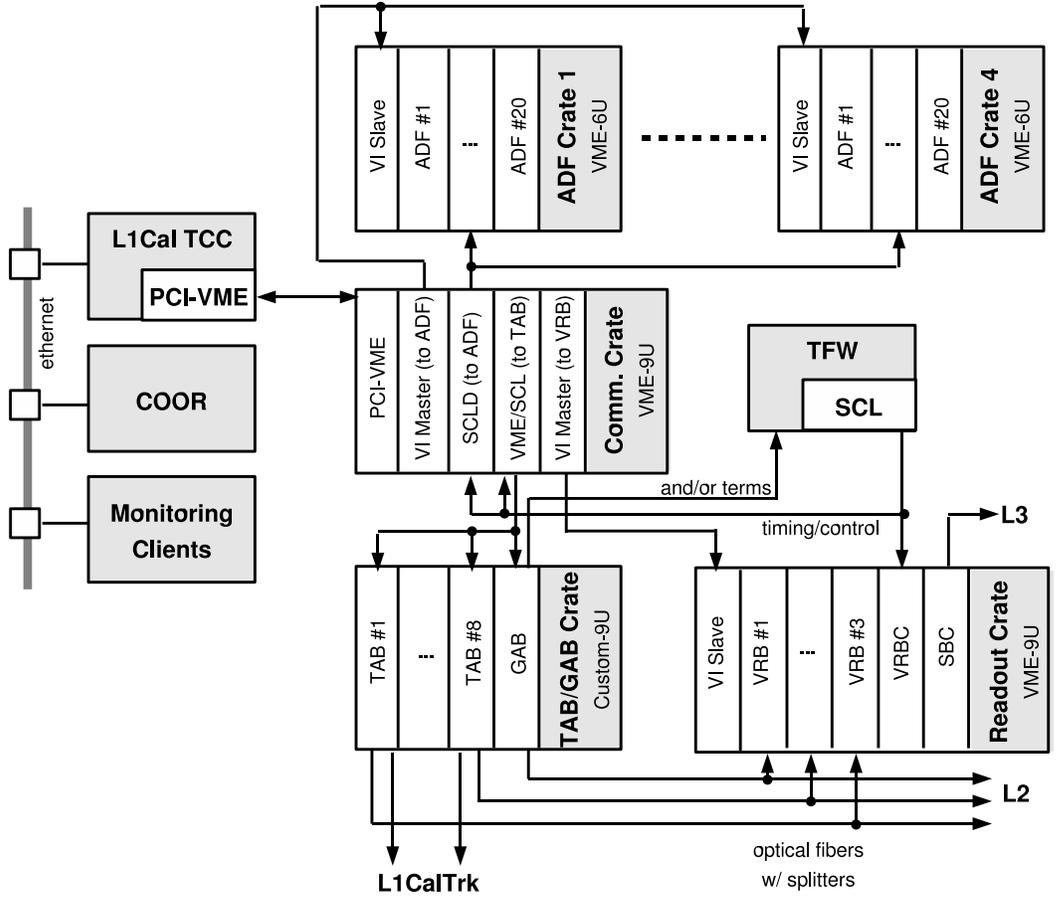}
  \caption{Communications in the L1Cal system.}
  \label{fig:communications}
\end{center}
\end{figure}

Most components of the \Dzero\ trigger and data acquisition system are
programmable.
The Online System allows this large set of resources and parameters to
be configured to support diverse operational modes --
broadly speaking, those used during proton-antiproton collisions in
the Tevatron ({\it physics modes}) and those used in the absence of
colliding beams ({\it calibration/testing modes}),
forming a large set of resources
and parameters needing to be configured before collecting data.

The L1Cal fits seamlessly into this Online System, with its
online control software hiding
the complexity of the underlying hardware, while making the run time
programming of the L1Cal Trigger accessible to all \Dzero\ users
in simple and logical terms. A diagram of the L1Cal, 
from an online data and control point of view, 
is shown in Fig.\ \ref{fig:communications}.
The main elements of L1Cal online control are listed below,
with those aspects specific to L1Cal described in more detail in the
following sections.
For more information on \Dzero -wide components see \cite{d0run2}.
\begin{itemize}
\item The Trigger Framework (TFW)
      delivers global \Dzero\ timing and control signals to the
      L1Cal and collects and/or terms from the GAB
      as described in Section \ref{sect:hardware}.
\item COOR \cite{d0run2},
      a central \Dzero\ application, coordinates all trigger
      configuration and programming requests.  Global trigger lists,
      containing requirements and parameters for all triggers used by
      the experiment, are specified using this application as are more
      specific trigger configurations (several of which may operate
      simultaneously) used for calibration and testing.
\item The L1Cal Trigger Control Computer (TCC), 
      a PC running the Linux operating system,
      provides a high
      level interface between COOR and the L1Cal hardware and allows
      independent expert control of the system.
\item The Communication Crate contains cards that provide an
      interface between the L1Cal custom hardware in the ADF and
      TAB/GAB crates, and the L1Cal TCC and SCL.
\item The L1Cal Readout Crate allows transmission of L1Cal data to
      the L3 trigger system.
\item Monitoring Clients, consisting of software that may run on a
      number of local or remote computers, display
      information useful for tracking L1Cal operational status.
\end{itemize}

%-------------------------------------------------------------
\subsection{L1Cal Control Path}
\label{sect:online-path}
The L1Cal Trigger Control Computer needs to access the 80 ADF cards in
their four 6U VME crates; 
the eight TAB and one GAB cards in one 9U custom crate;
and the readout support cards in one 9U VME crate.
It uses a commercial interface to the VME bus architecture --
the model 618 PCI-VME bus adapter \cite{sbs618}.
This adapter consists of one PCI module located in the TCC PC and
one VME card located in the Communication Crate, linked by an optical
cable pair.

To access the four ADF crates, the L1Cal system uses a set of
Vertical Interconnect (VI) modules built by
Fermilab \cite{vi}.
One VI Master Card is located in the Communication Crate,
and is connected to four VI Slaves, one in each ADF Crate.
The VI Master maps the VME A24 address space
of each remote ADF crate onto four contiguous segments of VME A32
addresses in the Communication Crate.
User software running on the L1Cal TCC generates VME A32/D16 cycles
in the Communication Crate, 
and A24/D16 in the ADF crate, via the VI Master-Slave interface.
The Communication crate also hosts one additional VI Master
to access a VI Slave located in the L1Cal Readout Crate.

As discussed in Section \ref{sect:vmescl},
VME transactions with the TAB/GAB crate are accomplished via the
VME/SCL card, housed in the Communication Crate.
User software running on L1Cal TCC
generates VME A24/D32 cycles to the VME/SCL, which in turn generates
a serialized transaction directly to the targeted TAB or GAB module.

%------------------------------------------------------------------
\subsection{L1Cal Control Software}
\label{sect:online-software}
The functionality required from the control software on the L1Cal TCC
is defined by three interfaces:
the COOR Interface,
the L1Cal Expert Interface, and
the Monitoring Interface.
The first two of these are used to configure and control 
L1Cal operations globally
(COOR) or locally when performing tests (Expert).
The Monitoring Interface collects monitoring information from the
hardware for use by Monitoring Clients (see Section
\ref{sect:monitor}). 

The L1Cal online code itself is divided into two parts:
the Trigger Control Software (TCS),
written in C++ and C,
where the main functionality of the
above three interfaces is implemented;
and the L1Cal Graphical User Interface (GUI),
written in Python \cite{python} with TkInter \cite{tkinter},
which allows experts to
interact directly with the TCS.

While the GUI normally runs on the L1Cal TCC computer,
it can also be launched from a different computer
located at \Dzero\ or at a remote institution.
Since it is a non-critical part of the control software,
the GUI does not need to run all the time, but
several instances of it can be 
started and stopped as desired, independently from the TCS.
Once started, an instance of the
GUI communicates with the TCS by exchanging XML (Extensible Markup
Language) \cite{xml} text strings.

For communications across each of its three interfaces
the TCS uses the ITC (Inter Task Communication) package
developed by \Dzero\ and based on the open-source ACE
(Adaptive Communication Environment) software \cite{ace}.
ITC provides high level management of client-server connections
where communication between separate processes,
which may be running on separate computers,
is dynamically buffered in message queues.
The TCS uses ITC to:
receive text commands from COOR and send
      acknowledgments back with the command completion status;
receive XML string commands from the GUI application
      and send XML strings back to the GUI; and
receive fixed format binary monitoring requests from the
      monitoring clients and send the requested fixed format binary
      block of data. 

%-------------------------------------------------------------------------
\subsection{Main Control Operations}
\label{sect:online-controlops}
Control operations in the L1Cal online software fall into three main
categories: configuration, initialization and run-time
programming. 
Configuration consists of loading pre-synthesized firmware into all
the FPGAs in the system. 
Initialization then
brings the system into a well-defined idle state.
During initialization, all control registers, geometric constants,
lookup tables, calibration parameters, etc. are overwritten with their
desired values.
It is also at this stage that problematic TTs are excluded from
consideration by programming their corresponding ADF cards to always
report zero \Et\ for the TTs in question.
The most IO intensive part of the initialization is in
the programming and verification of the 2,560 ADF \Et\ Lookup
Memories, which takes approximately five seconds.
After initialization, COOR performs the run time programming step,
where the specific meaning of each L1Cal trigger output signal (the
and/or terms) is defined. This involves loading \Et\ threshold values
and other algorithm parameters into the TABs as well as associating
combinations of objects and selection criteria in the GAB with
individual output and/or bits.
Once these tasks have been accomplished, the system runs
largely without external intervention, except for monitoring data
collection.

%==================================================================
% MONITORING
%==================================================================
\section{Managing Monitoring Information}
\label{sect:monitor}
The monitoring resources available in the ADF, TAB
and GAB cards are described in Sections 
\ref{sect:adf-proc} and \ref{sect:tabgab-diagnostic}.
This information is collected by the TCC Control Software 
and is made available to Monitoring Clients
via the Monitoring Interface
as outlined in
Section \ref{sect:online-software}.
During normal operation, monitoring data are collected approximately
every five seconds when the {\it Collect Status} qualifier is
asserted on the SCL along with \accept . 
If data flow has stopped,
monitoring data are still collected from the L1Cal, initiated by the
TCS, which times out after six seconds of
inactivity. 

Monitored data include the following.
\begin{itemize}
\item The ADF output \Et\ of all TTs 
      for all 36 active bunch crossings of the accelerator turn 
      containing the \accept\ for which
      the {\it Collect Status} signal is asserted.
\item The bunch crossing number within this turn that identifies
      the \accept .
\item The contents of all error and status registers in the TABs and
      GAB (associated with each SWA and Global chip on the TABs and
      with the LVDS and S30 chips on the GAB). 
      These registers indicate, among other information, synchronization
      errors on data transfer links, parity errors on each transfer,
      and bunch crossing number mismatches at various points in the
      TAB/GAB signal processing chain.
\end{itemize}

Monitoring information is displayed in the \Dzero\ control room and
remotely using Monitoring Client GUIs. This application,
written in Python \cite{python} with Tkinter \cite{tkinter},
requests and receives data from the TCS via calls to ITC.
It displays
average pedestal values and RMSs for each TT, to aid in the
identification of noisy or dead channels, as well as system status
information. 

Another tool for monitoring data quality in the control
room is a suite of ROOT-based \cite{root} software 
packages called {\it Examine}.  The L1Cal Examine
package receives a stream of data from L3 and displays
histograms of various quantities related to L1Cal performance,
including comparisons between L1Cal and calorimeter precision readout
estimates of TT energies.
Data
distributions can be compared directly to reference curves
provided on the plots, which can be obtained either from
an earlier sample of data or from simulation.

%==================================================================
% CALIBRATION
%==================================================================
\section{Calibration of the L1Cal}
\label{sect:calib}
Several methods are employed to ensure that the \Et\ of individual 
trigger towers, used
in the system, is correctly calibrated -- \ie , that one output count
corresponds to 0.25 GeV of \Et\ and that the zero-\Et\ baseline is set
to eight counts.

%------------------------------------------------------------------
\subsection{Online Pedestal Adjustment and Noise}
The most frequently used of these procedures is a tool, 
run as part of the TCS, 
which samples ADC-level data from the ADFs when no true energy is
expected to be deposited in the calorimeter. Based on this data, 
corrections to the DAC values used to set each channel's zero-energy
baseline are calculated and can be downloaded to the system.

This online pedestal adjustment is performed every few days because of
periodic pedestal shifts that occur in a small number of channels --
typically less than ten. These pedestal shifts arise because of
synchronous noise, with a period of 132 ns, observed in the system due
largely to pickup from the readout of other, nearby detector
systems. Although the amplitude of this noise varies from channel to
channel (it is largest in only a handful of TTs), 
its phase is stable over periods of several stores of particle beams
in the Tevatron,
which sets the timescale for pedestal readjustment.

%------------------------------------------------------------------
\subsection{Calorimeter Pulser}
The calorimeter pulser system \cite{d0run2}, which injects carefully
calibrated charge pulses onto the calorimeter preamps, is also used by
the L1Cal to aid in the identification of dead and noisy
channels. Special software compares \Et\ values observed in the ADFs
with expectations based on the pattern of preamps pulsed and the pulse
amplitudes used. Results are displayed graphically to allow easy
identification of problematic channels.
In addition to its utility in flagging bad channels, this system also
provides a quick way to check that the L1Cal signal path is properly
cabled. 

%------------------------------------------------------------------
\subsection{Offline Gain Calibration}

\begin{figure}
\begin{minipage}[t]{0.475\textwidth}
  \begin{center}
    \mygraphics{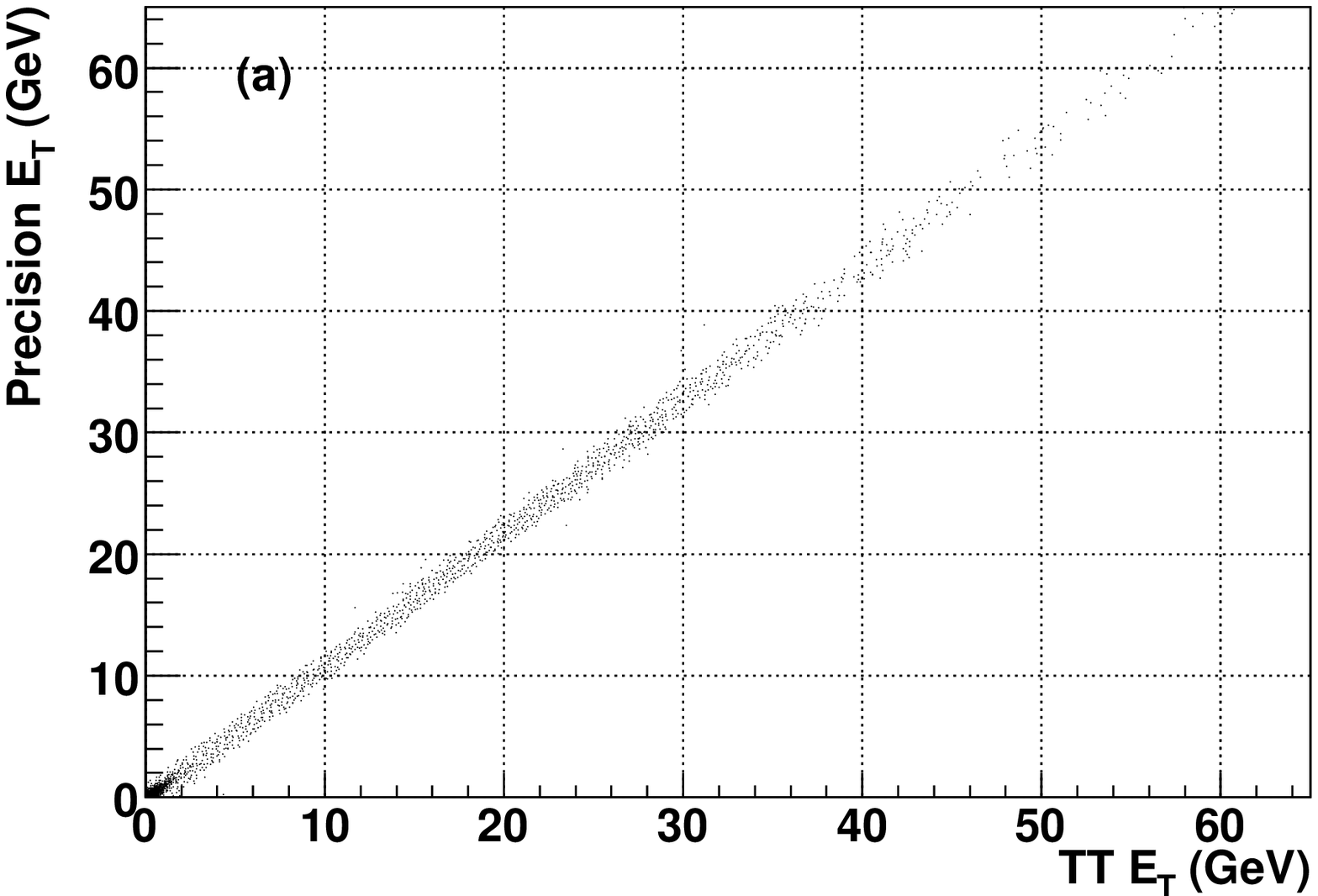}
  \end{center}
\end{minipage}
\hfill
\begin{minipage}[t]{0.475\textwidth}
  \begin{center}
    \mygraphics{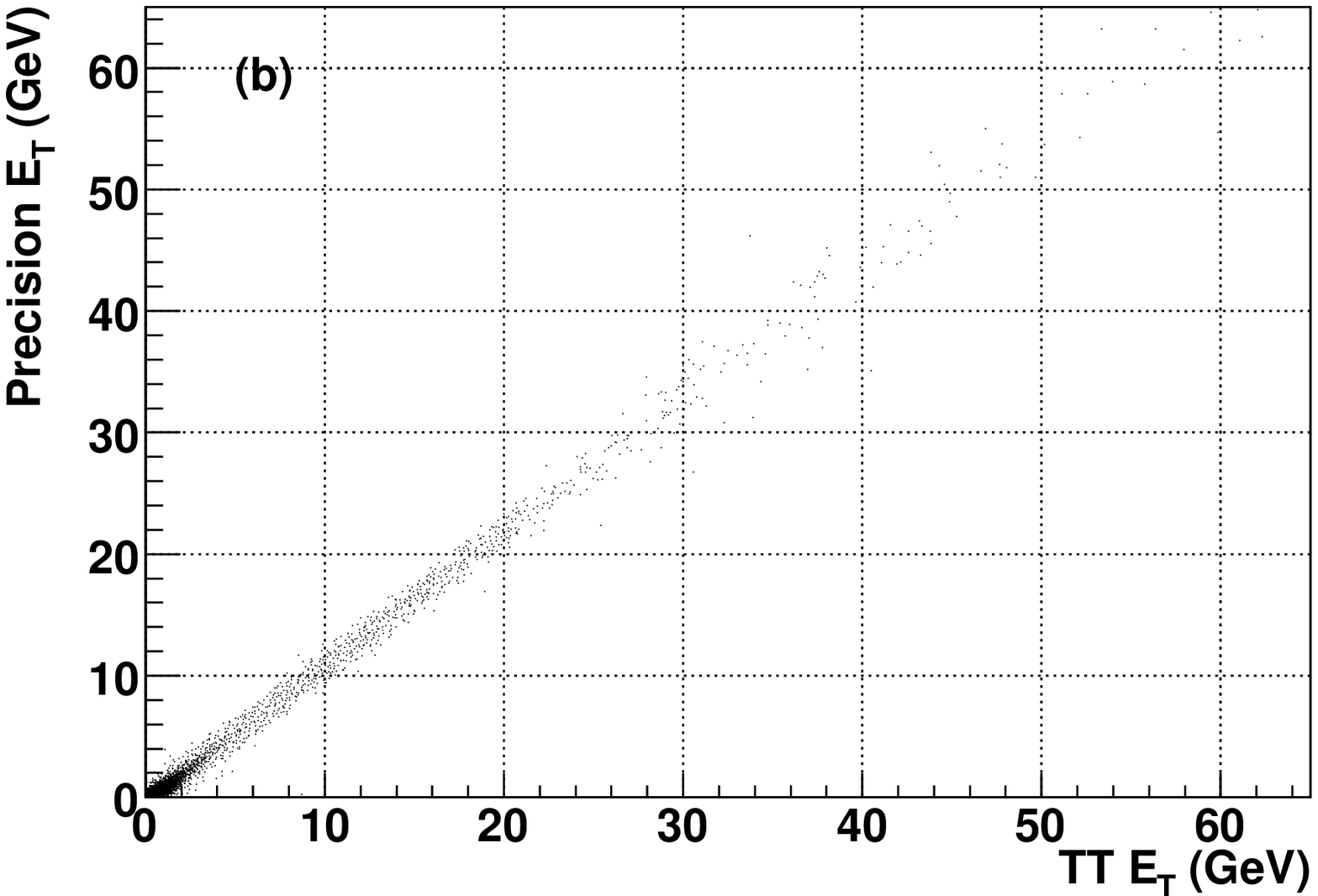}
  \end{center}
\end{minipage}
\begin{center}
  \caption{
    Precision versus TT \Et\ for one \EMtt\ (a) and one \HADtt\ (b)
    trigger tower. 
    The linear relationship with slope=1 indicates the good
    calibration of the tower.  The excursion away from an absolute
    correlation is an indication of the inherent noise of the system.}
  \label{fig:TTlin}
\end{center}
\end{figure}

The desired TT response of the L1Cal, 0.25 GeV per output count, is
determined by comparing offline TT \Et 's to the corresponding sums of
precision readout channels in the calorimeter,
which have already been calibrated against physics signals.
For this purpose, data taken during normal physics running of the
detector are used.
An example can be seen in Fig.\ \ref{fig:TTlin}. Gain calibration
constants, for use in the ADF \Et\ Lookup Memories, are derived from
the means of distributions of the ratio of TT to precision readout
channel sums for each \EMtt\ and \HADtt\ TT.

Gain coefficients derived in this way have been determined to be
stable to within $\sim$2\% over periods of months. Thus, this type of
calibration is normally performed only after extended Tevatron
shutdown periods.

%==================================================================
% RESULTS
%==================================================================
\section{Results}
\label{sect:results}

%------------------------------------------------------------------
\subsection{Run IIb Trigger List}
\label{sect:trig-list}
The trigger list for Run IIb was designed, 
with the help of the simulation tools described in Section
\ref{sect:sim}, 
to select all physics
processes of interest for the high luminosity running period, and to
run unprescaled at all instantaneous luminosities below \scinot{3}{32}
\instLunit .  
The entire Run IIb
L1 trigger menu normally produces an accept rate of up to 1800 Hz.
It includes a total of 256 and/or terms, of which 64 come from L1Cal,
falling into the following broad categories:
\begin{itemize}
\item one- two- and three-jet terms with higher jet
      multiplicity triggers requiring looser \Et\ cuts;
\item single- and di-EM terms without isolation requirements
      capturing high energy electrons;
\item single- and di-EM terms with isolation
      constraints
      (which currently consist of requiring that both the
      \EMtt /\HADtt\ and the EM-isolation ratios, described in
      Section \ref{sect:emalgo}, be greater than eight)
      designed for low energy electrons;  
\item tau terms, which select jets with three different isolation
      criteria;
\item topological terms, such as a jet with no other jet directly
      opposite to it in $\phi$, targeting specific signals that are
      difficult to trigger using the basic jet, EM and tau objects;
      and 
\item missing \Et\ terms.
\end{itemize}
These terms can be used individually,
or combined using logical ands, 
to form \Dzero\ L1 triggers in the TFW. 

%------------------------------------------------------------------
\subsection{Algorithm Performance and Rates}
\label{sect:perf}
\begin{figure}
\begin{center}
  \mygraphics{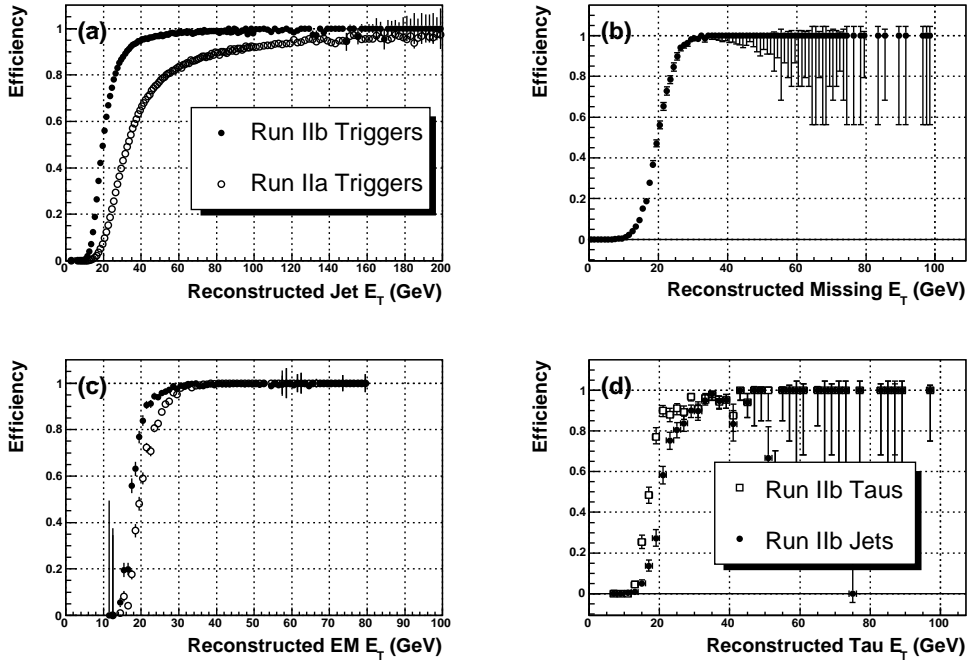}
  \caption{
    Trigger turn-on curves for 
    (a) single jet triggers using the Run IIa and Run IIb algorithms;
    (b) a Run IIb missing \Et\ trigger;
    (c) single EM object triggers using Run IIa and isolated Run IIb
    algorithms on offline selected \Zee\ events;
    and
    (d) a Run IIb single tau trigger compared with
    a Run IIb single jet trigger that runs at the same rate, using
    offline selected \Ztautau\ events.}
  \label{fig:perform} 
\end{center}
\end{figure}

L1Cal algorithm performance has been measured 
relative to unbiased offline
reconstruction of jets, electrons, taus, and missing \Et\
using runs taken at luminosities greater than \scinot{1}{32}
\instLunit ,
with a special ``low threshold'' trigger list
designed to minimize trigger bias in the data.
Some of these results are summarized in Fig.\ \ref{fig:perform}.
In Fig.\ \ref{fig:perform}(a), turn-on curves 
(efficiency vs. reconstructed jet \Et ) 
are shown for single jet triggers using Run IIb 
(with a jet-object \Et\ threshold of 15 GeV) 
and Run IIa
(requiring two TTs with \Et\ $>$ 5 GeV)
trigger algorithms.
The significantly steeper transition between low and high efficiency
for the Run IIb algorithm is evident here.
The turn-on curve for the Run IIb 20 GeV threshold missing \Et\
trigger is shown in Fig.\ \ref{fig:perform}(b).
The performance of this trigger is comparable to, or better than that
observed in Run IIa.
EM trigger performance is summarized 
in Fig.\
\ref{fig:perform}(c), 
for a sample of \Zee\ events, collected using unbiased
triggers. 
The plot shows
the efficiency vs. reconstructed EM object \Et\ 
for the logical OR of two separate trigger terms,
representative of trigger combinations used in electron-based
analyses at \Dzero .
The Run IIb terms used are
a single EM trigger term with a threshold of 19 GeV, or
an isolated single EM trigger with a threshold of 16 GeV;
while for Run IIa, the requirements are
a single TT with \EMtt\
\Et\ $>$ 16.5 GeV or two TTs with \Et\ $>$ 8.25 GeV.
Both of these triggers produce rates of 370--380 Hz at a luminosity of
\scinot{3}{32} \instLunit .
However, the Run IIb trigger combination
gives a sharper turn-on and allows for a lower
effective threshold, 
yielding a significantly higher efficiency for selecting \Zee\ decays
than that achievable using the Run IIa system.
Finally, results using the new Run IIb Tau algorithm are summarized in
Fig.\ \ref{fig:perform}(d).
In this plot, trigger turn-on curves are shown for 
single tau and single jet triggers
using a sample of
\Ztautau\ candidates, selected offline from
events collected using unbiased (muon) triggers.
The L1Cal tau algorithm allows lower object thresholds
to be used
(15 GeV taus compared to 20 GeV jets)
yielding higher signal selection efficiencies for the same trigger
rate.

Measured trigger rates using the new algorithms and trigger list are
consistent with those based on extrapolations of Run IIa data to Run
IIb instantaneous luminosities,
shown in Fig.\ \ref{fig:rates}.
As can be seen, 
the total trigger rate observed using the new Run IIb list,
to which L1Cal contributes more than 50\% of the events,
fits into the
bandwidth limitations of the experiment.
A Run IIa trigger list, designed to give the same selection efficiency
as the Run IIb list above,
would have exceeded these limits by a factor of two or more.

%==================================================================
% CONCLUSIONS
%==================================================================
\section{Conclusions}
\label{sect:conclusion}
The new D0 Run IIb L1Cal trigger system was designed to cope efficiently
with the highest instantaneous luminosities foreseen during the Run
IIb operating period of the Tevatron at Fermilab.
To accomplish this goal clustering algorithms have been developed
using a novel hardware architecture that uses bit-serial data
transmission and arithmetic to produce a compact, cost-effective
system built using commercially available FPGAs.
Although data transmission rates in the system approach one tera-bit
per second, the system has been remarkably stable since
it began to operate at the beginning of Run IIb.

With the Tevatron regularly producing instantaneous luminosities in
excess of \scinot{2}{32} \instLunit , the new trigger system has been
tested extensively at its design limits. So far it has performed
exceptionally well, achieving background rejection factors sufficient to fit
within the bandwidth limitations of the experiment while retaining the
same or better efficiencies as observed in Run IIa 
for interesting physics processes.

%==================================================================
% ACKNOWLEDGMENTS
%==================================================================
\section{Acknowledgments}
\label{sect:acknowledgments}
We gratefully acknowledge the guidance of George Ginther, Jon
Kotcher, Vivian O'Dell, and Paul Padley as managers of the Run IIb
project,
as well as the technical advice of Dean Schamberger.
We would also like to thank 
Samuel Calvet, Kayle DeVaughan, Ken Herner, Marc Hohlfeld, Bertrand
Martin, and Thomas Millet
for analyzing L1Cal data and
for producing the performance plots shown in this paper. 
Finally, we thank the staffs at Fermilab and the collaborating
institutions, and acknowledge support from the DOE and NSF (USA);
CEA (France); and the CRC Program, CFI and NSERC (Canada).

%==================================================================
% BIBLIOGRAPHY
%==================================================================


\begin{thebibliography}{00}
\bibitem{etaphi}
  The right-handed \Dzero\ coordinate system is defined 
  with the $z$-axis in the direction of the proton beam,
  with $\phi$ measuring the
  azimuthal angle in the plane transverse to the beam direction,
  with $\theta$ measuring the polar angle,
  and with the pseudo-rapidity, $\eta = -\ln [ \tan (\theta / 2) ]$.
\bibitem{caltrkdata}
  Because of the extra latency involved in the divide operation
  necessary for the calculation of the Tau algorithm isolation, this
  information is not available in time for transmission to the
  L1CalTrk system.
\end{thebibliography}
\end{document}